\documentclass[acmtog]{acmart}
\hypersetup{draft}
\usepackage[english]{babel}
\usepackage{booktabs} 
\usepackage{wrapfig}
\usepackage{subcaption}
\usepackage{stackengine}
\usepackage[normalem]{ulem}
\usepackage{algorithm}
\usepackage{algpseudocode}
\usepackage{adjustbox}
\usepackage{tikz}
\usepackage{array}

\acmPrice{15}

\setcopyright{acmcopyright}

\setcitestyle{square}

\usepackage{xcolor} 
\usepackage{graphicx}

\renewcommand{\vec}[1]{\mathbf{#1}}
\newcommand{\mat}[1]{\mathbf{#1}}

\newcolumntype{?}{!{\vrule width 1pt}}

\newcommand{\todo}[1]{}

\newcommand\hide[1]{}

\newcommand\revise[1]{{#1}}

\definecolor{bkcolor}{RGB}{210,10,210}
\definecolor{vcacolor}{RGB}{123,50,210}
\definecolor{barbaracolor}{RGB}{246,150,70}

\newcommand\eg{e.g.,~}
\newcommand\Fig[1]{Figure~\ref{fig:#1}}
\newcommand\Sec[1]{Section~\ref{sec:#1}}
\newcommand\Eq[1]{Equation~(\ref{eq:#1})}
\newcommand\Algo[1]{Algorithm~(\ref{algo:#1})}
\newcommand\Tab[1]{Table~(\ref{tab:#1})}

\newcommand\twoD{2D}
\newcommand\threeD{3D}

\DeclareMathOperator*{\argmin}{arg\,min}

\newcommand\tablefont[1]{\small{#1}} 
\newcommand\imagefont[1]{\footnotesize{#1}}

\begin{document}

\acmJournal{TOG}
\acmYear{2019}\acmVolume{38}\acmNumber{6}\acmArticle{188}\acmMonth{11} \acmDOI{10.1145/3355089.3356560}

\title{Transport-Based Neural Style Transfer for Smoke Simulations}

\author{Byungsoo Kim}
\affiliation{%
   \institution{ETH Zurich}}
\email{kimby@inf.ethz.ch}

\author{Vinicius C. Azevedo}
\affiliation{%
  \institution{ETH Zurich}}
\email{vinicius.azevedo@inf.ethz.ch}

\author{Markus Gross}
\affiliation{%
   \institution{ETH Zurich}}
\email{grossm@inf.ethz.ch}

\author{Barbara Solenthaler}
\affiliation{%
   \institution{ETH Zurich}}
\email{solenthaler@inf.ethz.ch}

\acmSubmissionID{}
\renewcommand{\shortauthors}{B. Kim, V. C. Azevedo, M. Gross, B. Solenthaler}

\begin{teaserfigure}
	\hspace{65px}
	\includegraphics[width=0.75\textwidth]{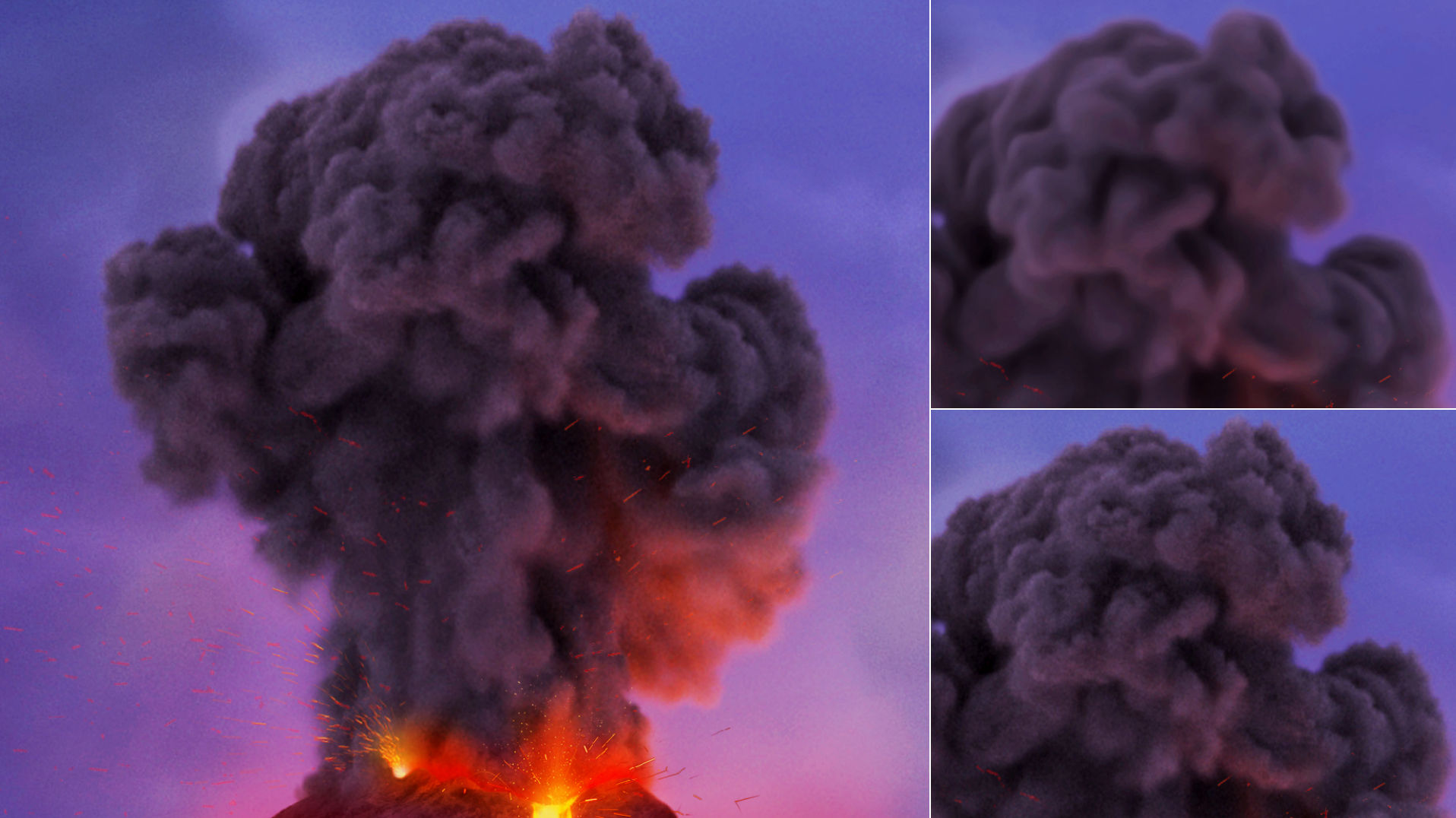}
	\hspace{-178px} 
    \includegraphics[trim=0px 0px 0px 0px, clip, width=0.07\textwidth]{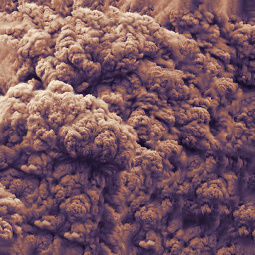}   
  	\hspace{-75px}
    \includegraphics[trim=0px 0px 0px 0px, clip, width=0.07\textwidth]{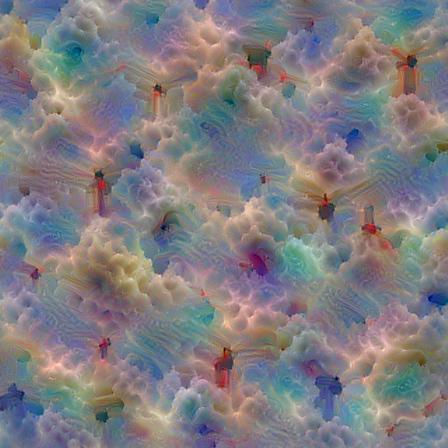}
	\caption{Volcanic smoke simulation. Left: stylized output by our transport-based neural style transfer; right: close-up views of low-resolution base input and ours \revise{with a cloud motif$^0$ and a smoke exemplar \textcopyright Richard Roscoe via~\cite{jamrivska2015lazyfluids}.}}
	\label{fig:teaser}
\end{teaserfigure}

\begin{abstract}
Artistically controlling fluids has always been a challenging task. Optimization techniques rely on approximating simulation states towards target velocity or density field configurations, which are often handcrafted by artists to indirectly control smoke dynamics. Patch synthesis techniques transfer image textures or simulation features to a target flow field. However, these are either limited to adding structural patterns or augmenting coarse flows with turbulent structures, and hence cannot capture the full spectrum of different styles and semantically complex structures. In this paper, we propose the first Transport-based Neural Style Transfer (TNST) algorithm for volumetric smoke data. Our method is able to transfer features from natural images to smoke simulations, enabling general content-aware manipulations ranging from simple patterns to intricate motifs.
%
%
The proposed algorithm is physically inspired, since it computes the density transport from a source input smoke to a desired target configuration. Our transport-based approach allows direct control over the divergence of the stylization velocity field by optimizing incompressible and irrotational potentials that transport smoke towards stylization. Temporal consistency is ensured by transporting and aligning subsequent stylized velocities, and 3D reconstructions are computed by seamlessly merging stylizations from different camera viewpoints.
\end{abstract}





\maketitle
\footnotetext{\url{http://storage.googleapis.com/deepdream/visualz/tensorflow_inception/index.html}}

\section{Introduction}
\label{sec:Introduction}

\begin{figure*}[t!]
    \centering
        \includegraphics[trim=250px 0px 250px 0px, clip,width=0.14\textwidth]{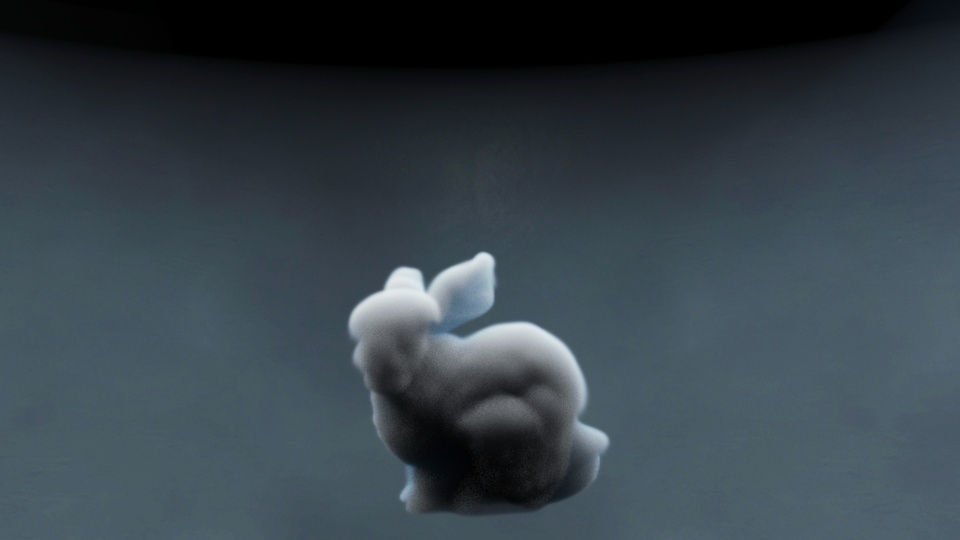}\hspace{-2px}
        \includegraphics[trim=250px 0px 250px 0px, clip,width=0.14\textwidth]{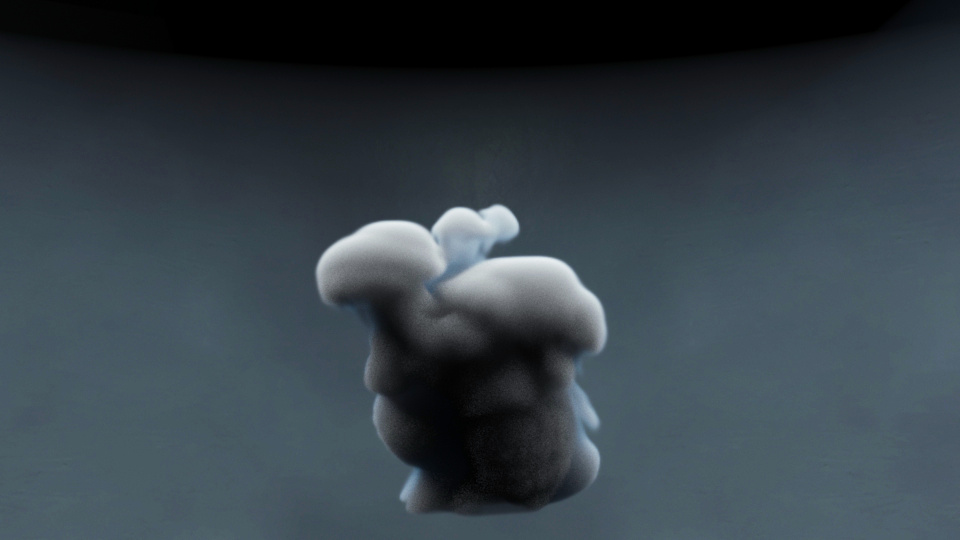}\hspace{-2px}
        \includegraphics[trim=250px 0px 250px 0px, clip,width=0.14\textwidth]{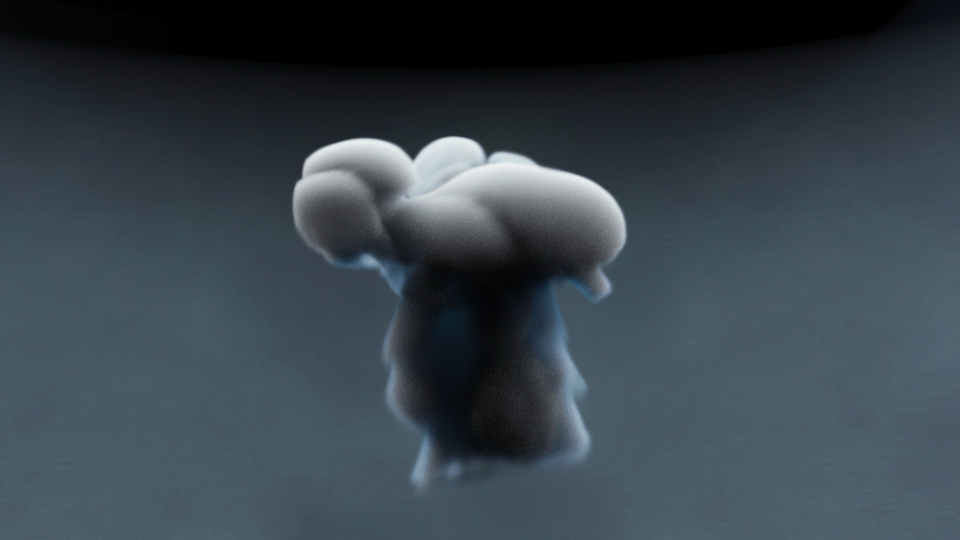}\hspace{-2px}
        \includegraphics[trim=250px 0px 250px 0px, clip,width=0.14\textwidth]{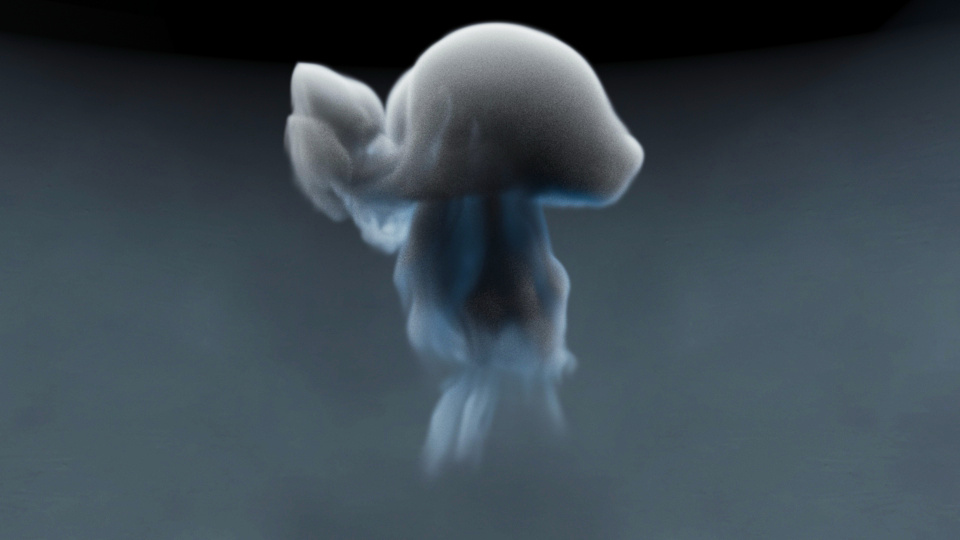}\hspace{-2px}
        \includegraphics[trim=250px 0px 250px 0px, clip,width=0.14\textwidth]{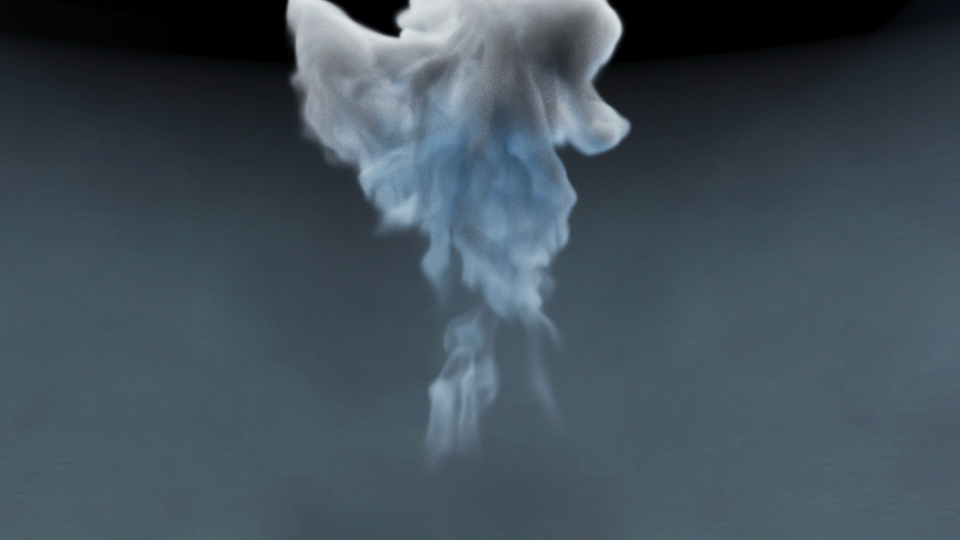}
        \\ \vspace{-1px}
        \includegraphics[trim=250px 0px 250px 0px, clip,width=0.14\textwidth]{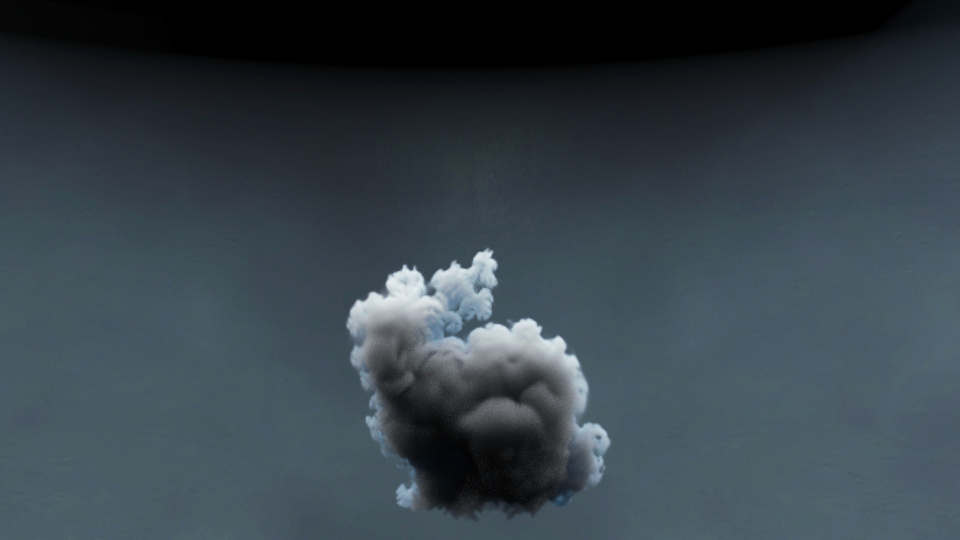}\hspace{-2px}
        \includegraphics[trim=250px 0px 250px 0px, clip,width=0.14\textwidth]{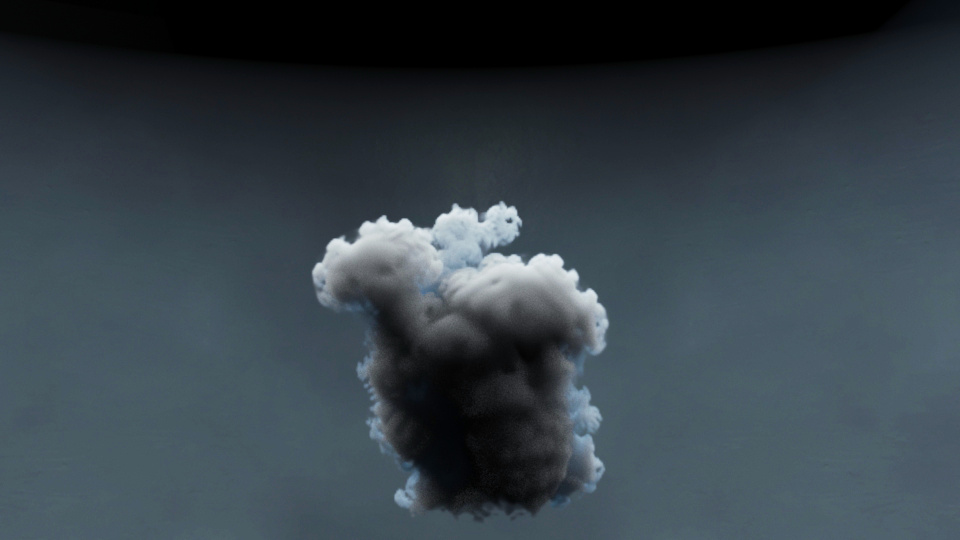}\hspace{-2px}
        \includegraphics[trim=250px 0px 250px 0px, clip,width=0.14\textwidth]{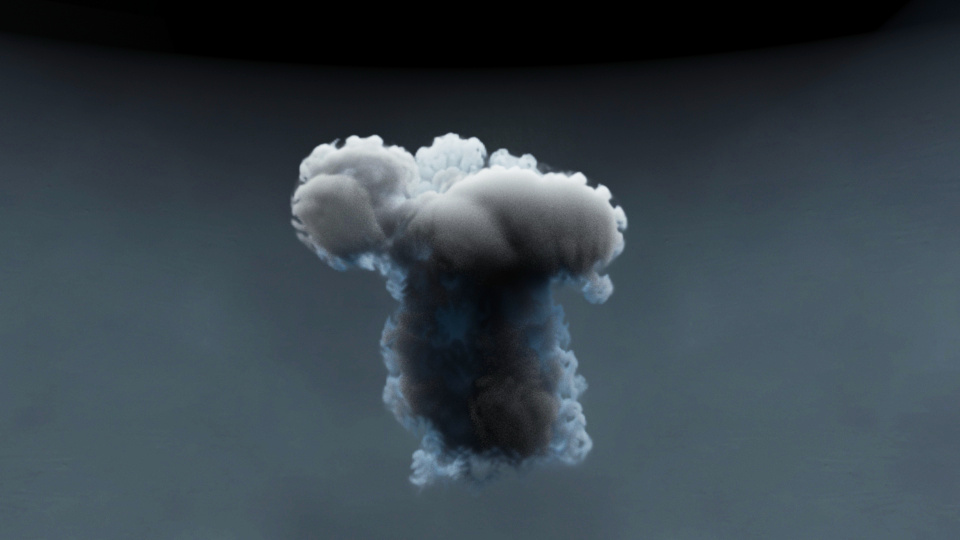}\hspace{-2px}
        \includegraphics[trim=250px 0px 250px 0px, clip,width=0.14\textwidth]{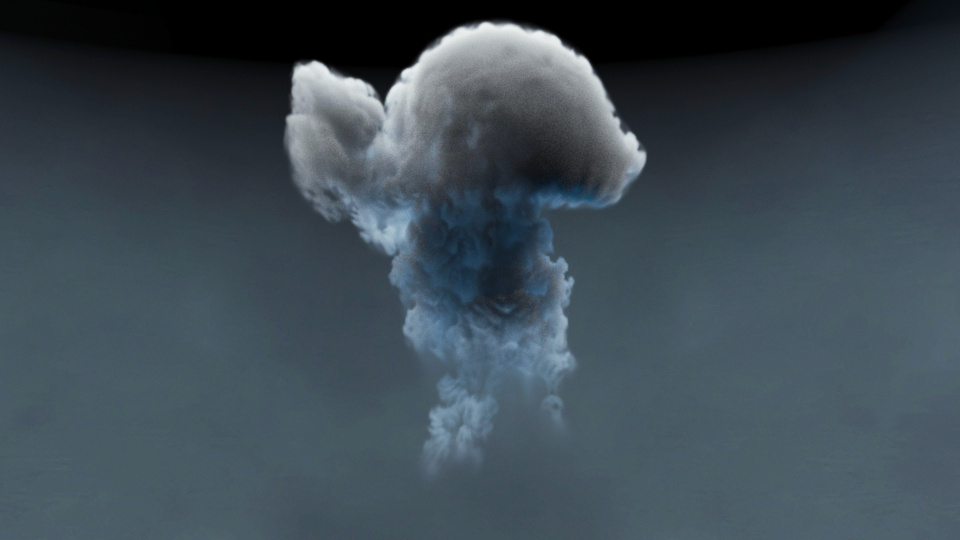}\hspace{-2px}
        \includegraphics[trim=250px 0px 250px 0px, clip,width=0.14\textwidth]{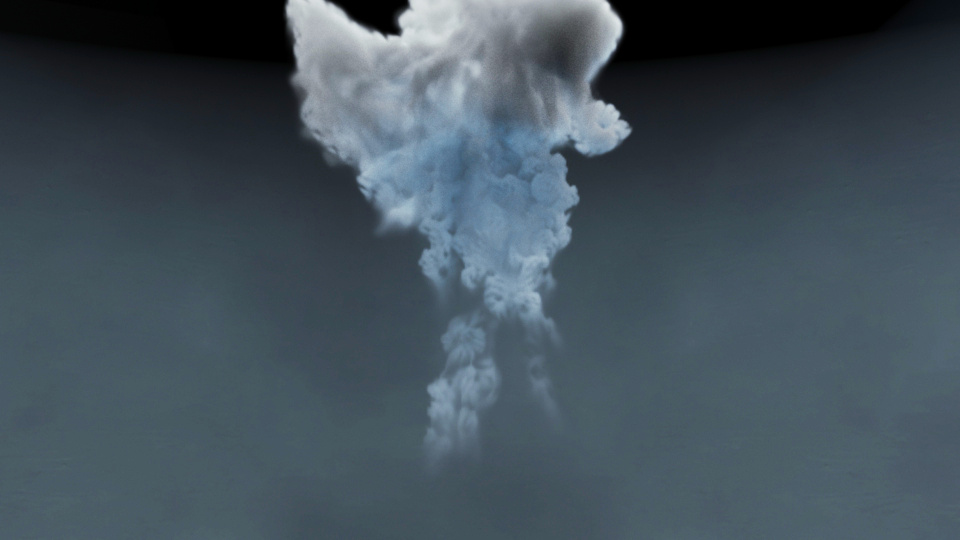}
        \hspace{-30px}
    	\includegraphics[trim=0px 0px 0px 0px, clip, width=0.05\textwidth]{img/style/volcano}
        \\ \vspace{-1px}
        \includegraphics[trim=250px 0px 280px 0px, clip,width=0.14\textwidth]{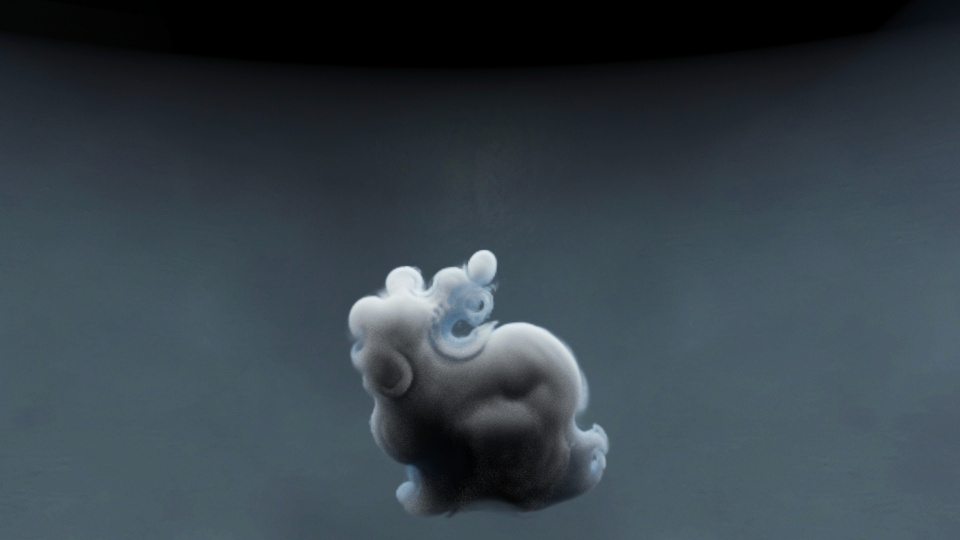}\hspace{-2px}
        \includegraphics[trim=250px 0px 280px 0px, clip,width=0.14\textwidth]{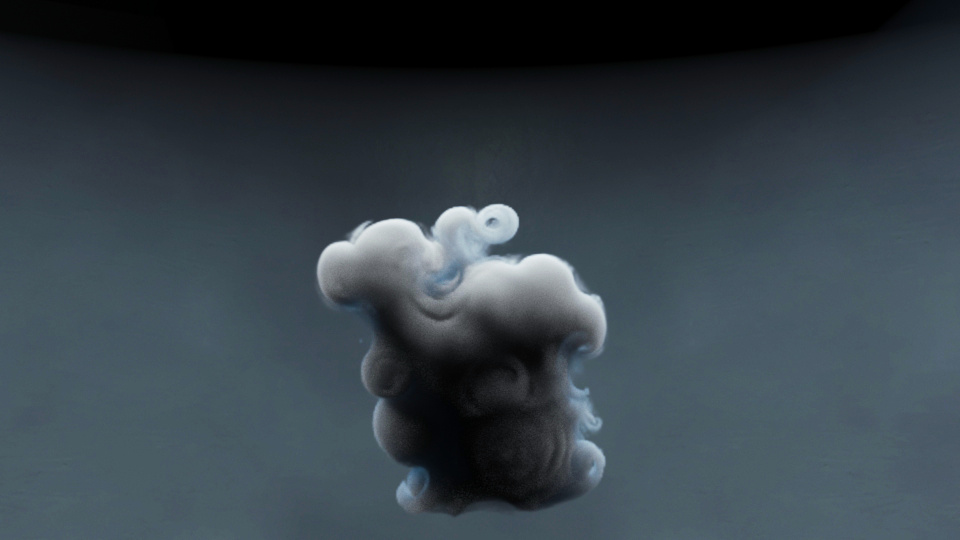}\hspace{-2px}
        \includegraphics[trim=250px 0px 280px 0px, clip,width=0.14\textwidth]{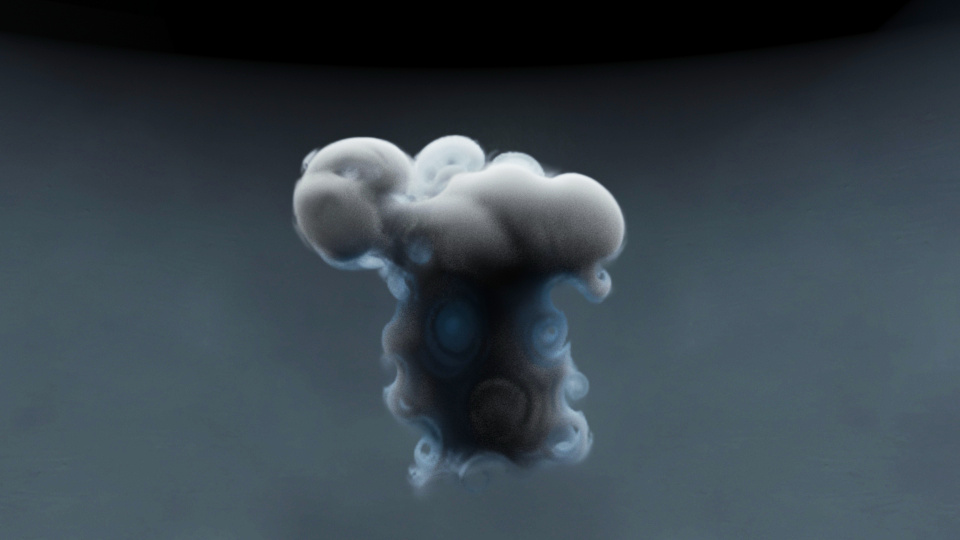}\hspace{-2px}
        \includegraphics[trim=250px 0px 280px 0px, clip,width=0.14\textwidth]{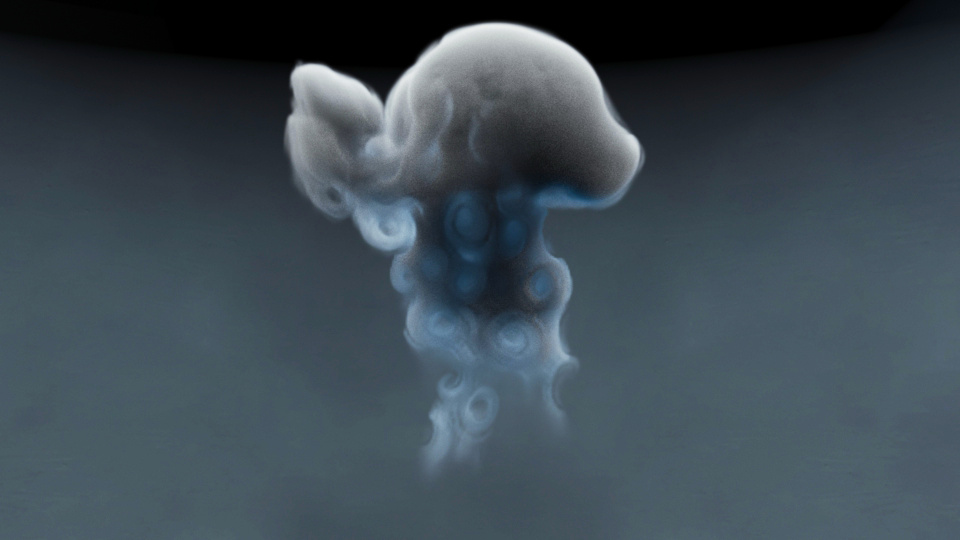}\hspace{-2px}
        \includegraphics[trim=250px 0px 280px 0px, clip,width=0.14\textwidth]{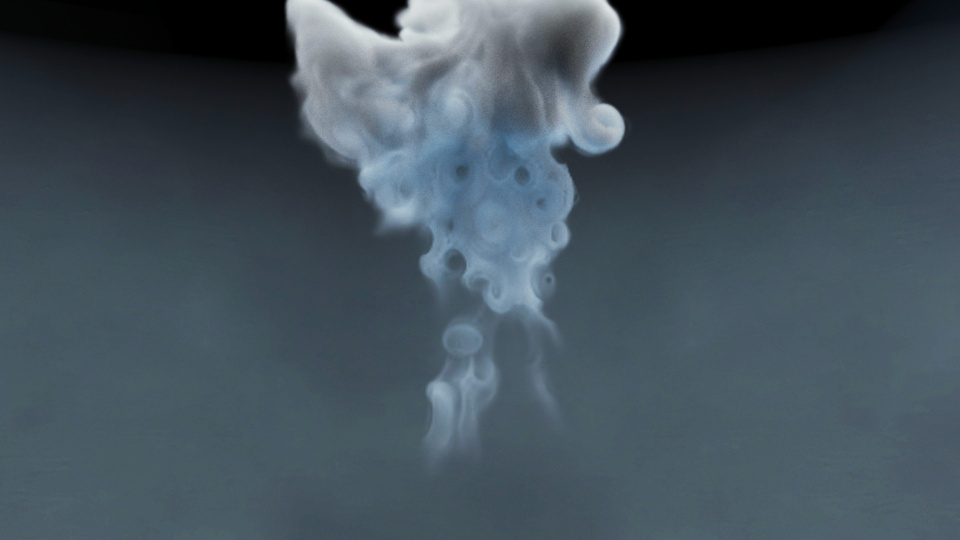}
        \hspace{-30px}
    	\includegraphics[trim=0px 0px 0px 0px, clip, width=0.05\textwidth]{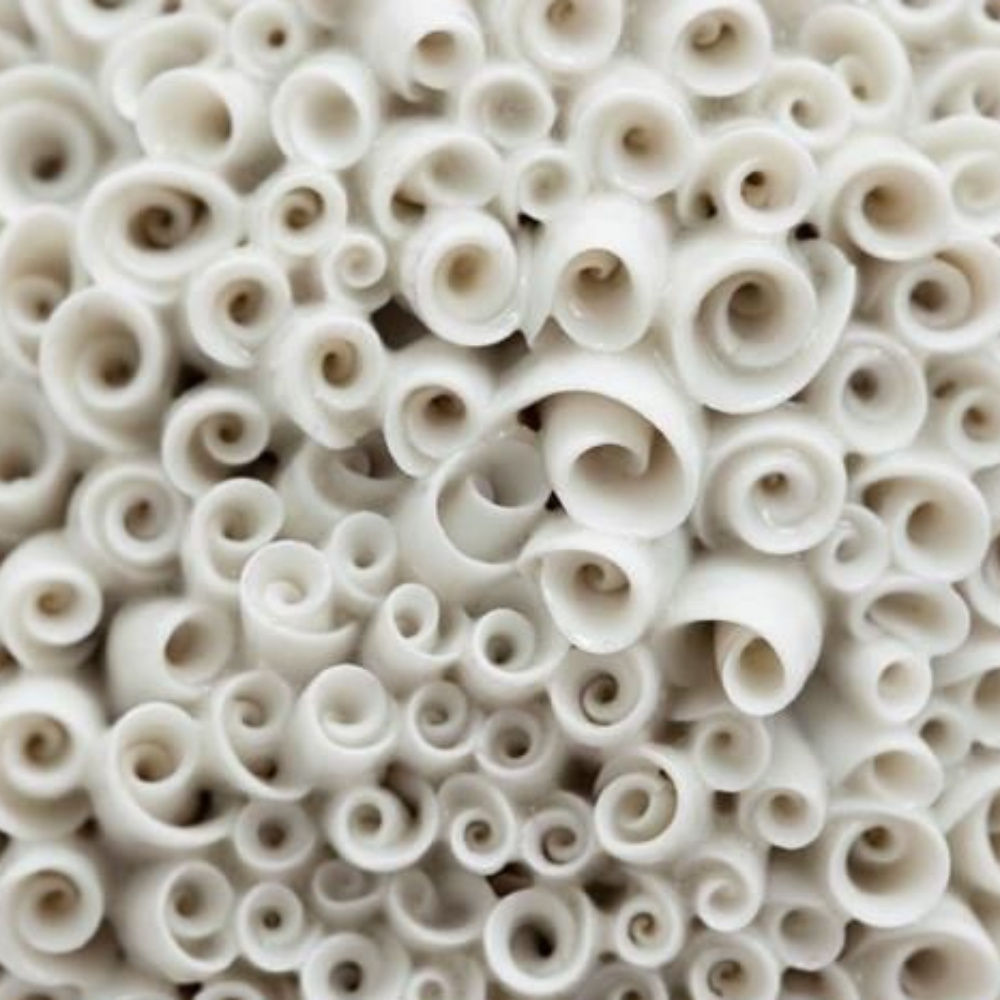}
        \\
      \caption{Frames of a style transfer smoke example: base simulation (top), stylized output with volcano (middle\revise{, \textcopyright Richard Roscoe}) and spiral\protect\footnotemark~(bottom) images.}
    \label{fig:sequence}
    \vspace{-5px}
\end{figure*}

Physically-based fluid simulations have become an essential part in digital content production. Due to the complexity of the underlying mathematical models, the process of manipulating fluids for simultaneously achieving controllable and realistic behavior, however, is tedious and time-consuming. Previous approaches \cite{Treuille2003, Inglis2016} relied on optimization techniques to generate artificial forces in a flow solver to match user-designed keyframes for smoke \cite{Treuille2003} and liquid animations \cite{Nielsen2011}. Optimization techniques are computationally challenging since the space of control forces is naturally large \cite{Pan2016}, limiting the applicability of these methods to relatively coarse grid resolutions. Moreover, since these techniques rely on manually crafted keyframes, artistic manipulation is restricted to reproducing given \threeD~shapes in their entirety, while modification of small to medium scale flow features is not easily attainable.
%

Post-processing methods for fluids aim at enabling detailed feature control by patch-based texture and velocity synthesis. While current patch-based techniques focus on controlling structural patterns~\cite{ma2009motion, jamrivska2015lazyfluids}, they are limited to \twoD~flows. Velocity synthesis approaches allow augmentation of coarse simulations with turbulent structures~\cite{kim2008wavelet, sato2018example}, but cannot capture the full spectrum of different styles and complex semantics. Ideally, to support artistic manipulations of flow data, post-processing methods should enable multi-level control of flow features with automatic instantiation of patterns.

Inspired by Neural Style Transfer (NST) methods for images~\cite{gatys2015neural} and meshes~\cite{kato2018neural, liu2018paparazzi}, we propose a novel method to synthesize semantic structures onto volumetric flow data by taking advantage of the simple yet powerful machinery developed for image editing. We modify \threeD~density fields by combining individual \twoD~stylizations from multiple views, which are synthesized by matching features of a pre-trained Convolutional Neural Network (CNN). Since the CNN is trained for image classification tasks, a vast library of patterns and class semantics is available, enabling novel content-aware flow manipulations that range from transferring low (edges and patterns) to high (complex structures and shapes) level features from images to smoke simulations (Figures \ref{fig:teaser}, \ref{fig:semanticTransfer} and \ref{fig:styleTransfer}). In this way, our method allows for automatic instantiation of structures in flow regions that naturally share features with a given target pattern or semantic class.

Crucially, and in contrast to existing NST methods, our style transfer algorithm is physically inspired. It computes a velocity field that stylizes a smoke density with an input target style, yielding results that naturally model the underlying transport phenomena. To improve temporal consistency, we propose a method which aligns stylization velocities from adjacent frames, enabling the control of how smoothly stylized structures change in time. To handle volumetric smoke stylization, multiple stylized \twoD~views are seamlessly combined into a \threeD~representation, resulting in coherent stylized smoke structures from arbitrary camera viewpoints. The proposed method is end-to-end differentiable and it can be readily optimized by gradient descent approaches. This is enabled by a novel volumetric differentiable smoke renderer, which is tailored for stylization purposes. Our results demonstrate that our method captures a wide spectrum of different styles and high-level semantics, and hence can be used to transfer patterns and regular structures, turbulence effects, shapes and artistic styles onto existing simulations.
\footnotetext{\url{http://ardezart.com/?attachment_id=252}}

\section{Related Work}
\label{sec:RelatedWork}


\begin{figure*}[t!]
    \centering
    \includegraphics[trim=325px 0px 325px 225px, clip, width=0.14\textwidth]{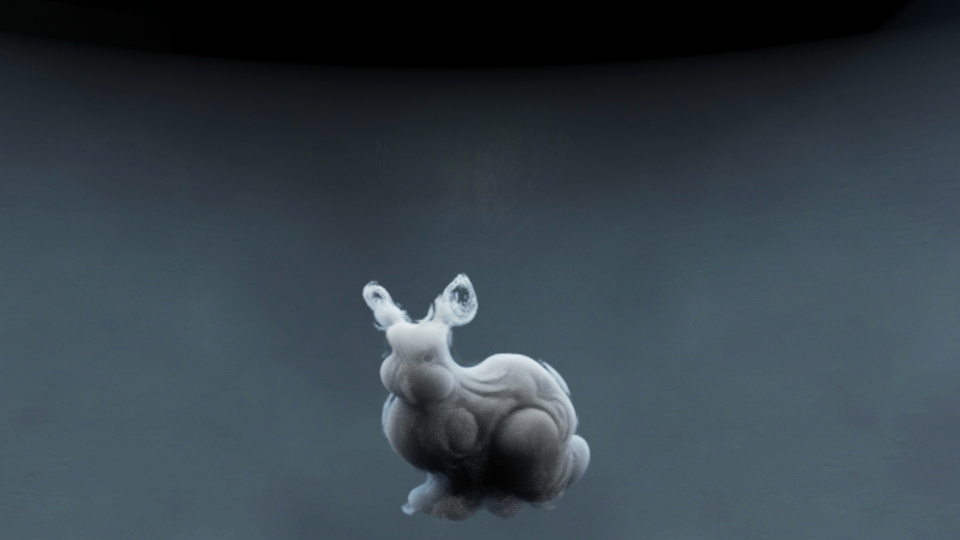}\hspace{-2px}
    \includegraphics[trim=325px 0px 325px 225px, clip, width=0.14\textwidth]{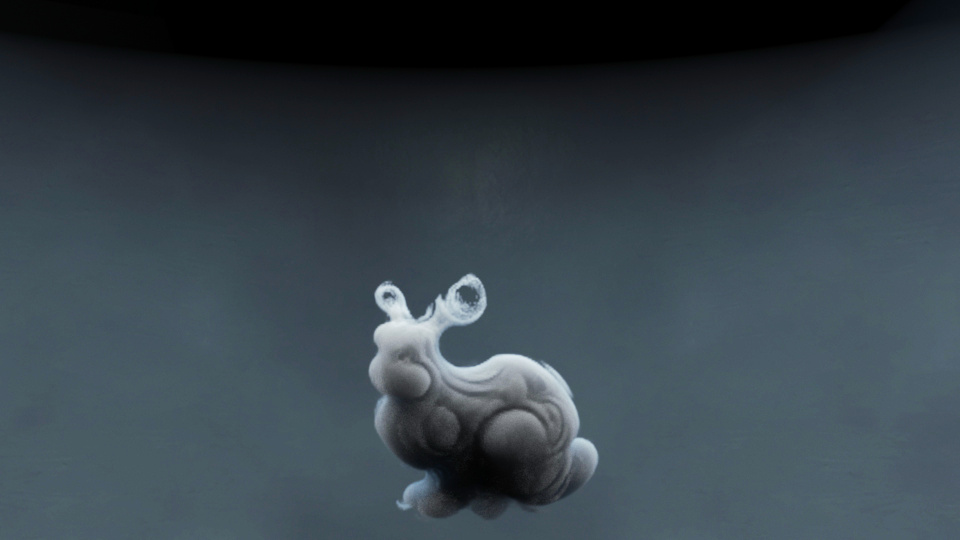}\hspace{-2px}
    \includegraphics[trim=325px 0px 325px 225px, clip, width=0.14\textwidth]{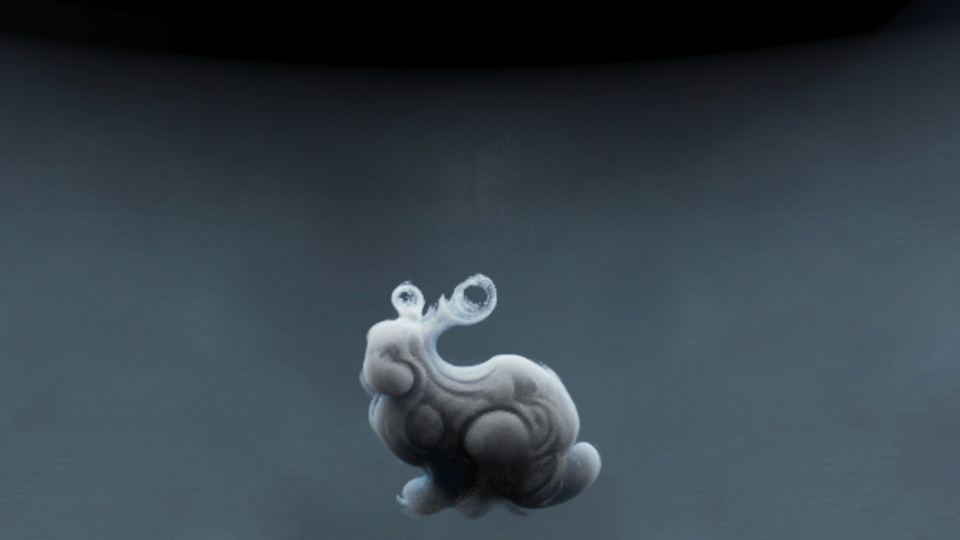}\hspace{-2px}
    \includegraphics[trim=325px 0px 325px 225px, clip, width=0.14\textwidth]{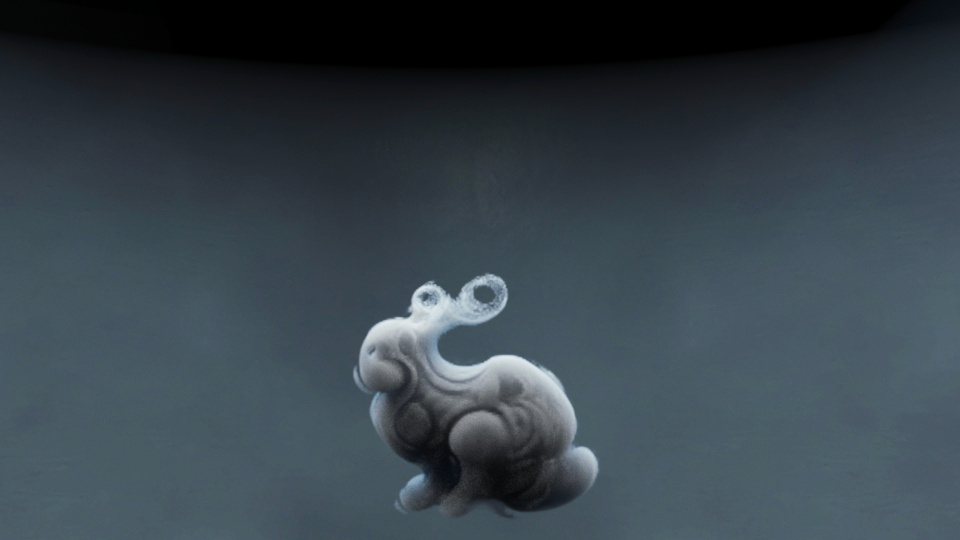}\hspace{-2px}
    \includegraphics[trim=325px 0px 325px 225px, clip, width=0.14\textwidth]{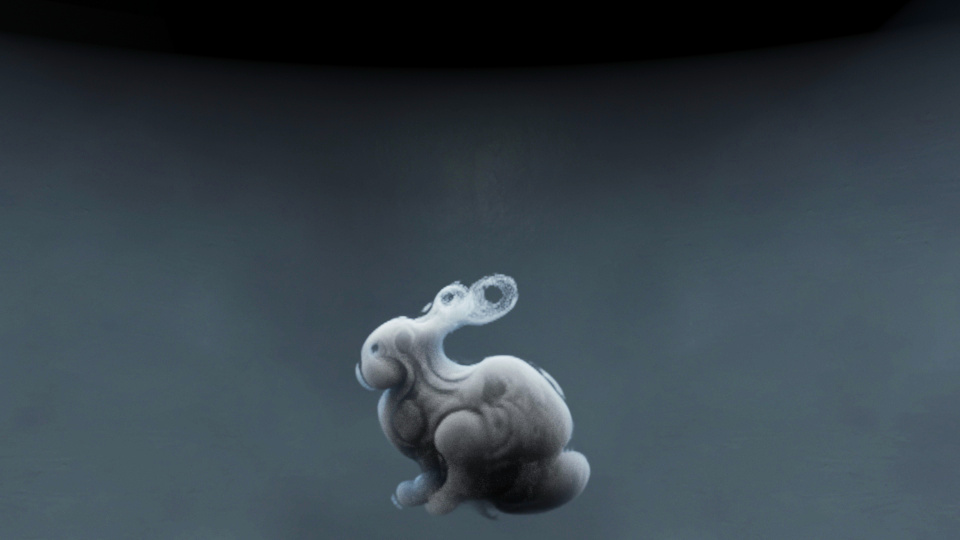}
    \hspace{-30px}
    \includegraphics[trim=0px 0px 0px 0px, clip, width=0.05\textwidth]{img/style/pattern1}
    \caption{Stanford Bunny shaped smoke stylized with spiral patterns$^1$ for multiple views \revise{([-30$^{\circ}$,30$^{\circ}$], every 15 degrees)}. Our method focuses the instantiation of patterns on smoke regions that share similarities with the target motif. Additionally, augmented flow structures change smoothly when the camera moves around the object.}
    \label{fig:multiView}
    \vspace{-5px}
\end{figure*}

\emph{Patch-based Appearance Transfer} methods change the appearance of a source image or texture to a target by matching small compact regions called patches or neighborhoods. Kwatra et al. \shortcite{Kwatra2005} employ local similarity measures to optimize an energy-based formulation, enabling the animation of texture patches by flow fields. Their approach was extended to liquid surfaces \cite{Kwatra2006, Bargteil2006}, and further improved by modifying the underlying texture based on visually salient features of the liquid mesh \cite{Narain2007}. Bousseau et al. \shortcite{Bousseau2007} proposed a bidirectional advection scheme to reduce local patch distortions in a video watercolorization setup. Regenerative morphing and image melding techniques were combined with patch-based tracking to produce in-betweens for artist-stylized keyframes \cite{browning2014stylized}. Jamri\v{s}ka et al. \shortcite{jamrivska2015lazyfluids} improved the temporal coherence aspect of previous energy-based formulations by reducing the wash-out effects that appear when textures are advected during long periods. Although patch-based appearance transfer methods were successful in synthesizing temporally coherent textures for flow animations, these are limited to \twoD~setups. For a broad discussion of patch-based texture synthesis works we refer to \cite{Barnes2017}.

\begin{figure*}[t!]
    \centering
    \includegraphics[trim=0px 0px 0px 0px, clip, width=0.75\textwidth]{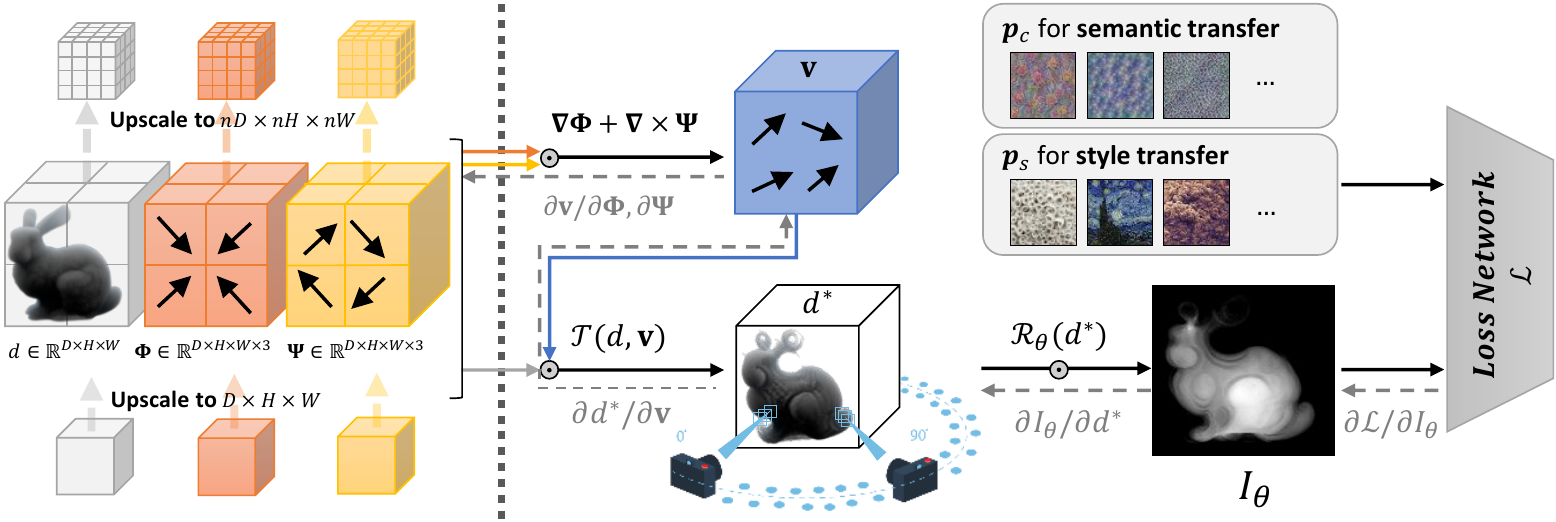}
    \caption{\revise{Pipeline of our TNST method. On the left, three input fields $d, \Phi, \vec{\Psi}$ for the stylization algorithm are shown. $\Phi, \vec{\Psi}$ are iteratively updated during the optimization, while the stylized density output is represented by $d^*$. All fields are firstly downsampled and stylized on their coarser representations, so features can be enhanced through larger regions of the smoke. Cubic upsampling is performed on $d, \Phi, \vec{\Psi}$ and the stylization runs again on a finer resolution; this process repeats until the specified resolution is matched. The right side of the diagram illustrates how our optimization works. The black arrows show the direction of a feed-forward pass from potentials $\Phi, \vec{\Psi}$ to the loss network, and the gray arrows represent the backpropagation path for computing gradients.}}
    \label{fig:pipeline}
    \vspace{-5px}
\end{figure*}

\vspace{1.5mm}
\emph{Velocity Synthesis} methods augment flow simulations with detailed flow fields to produce a desired effect. Due to the inability of numerical solvers to capture different energy scales of flow phenomena, sub-grid turbulence~\cite{kim2008wavelet, Schechter2008} was modelled for increased realism. This was later extended to model turbulence in the wake of solid boundaries \cite{Pfaff2009} and liquid surfaces~\cite{kim2013closest}. Sato et al. \shortcite{sato2018example} transferred turbulence data from a source to a target scene similarly to patch-based appearance transfer methods: the target simulation is subdivided into smaller patches which are matched to the ones of the source simulation. Patterns are matched by the combination of patchwise weighted L2 distance functions on velocity and density data, performed in a two-level search. However, their method is limited to turbulent features only, and more general style transfer between distinct simulations is not demonstrated. Ma et al. \shortcite{ma2009motion} synthesized velocity fields with example-based textures for artistic manipulations, but their method is limited to simple \twoD~patterns.


\vspace{1.5mm}
\emph{Fluid Control} aims to define the overall shape and behavior through user-specified keyframes or reference images. Optimal \cite{Treuille2003, Mcnamara2004} and proportional-derivative \cite{Fattal2004, Shi2005} controllers define a set of forces that guide fluid simulation states to desired configurations. These methods were extended to match simulations from different resolutions \cite{Nielsen2009}, guide fluids \cite{Rasmussen2004, Nielsen2011}, volume-preserving morphing \cite{Raveendran2012}, improve performance \cite{Pan2016}, and model more accurate boundary conditions while distinguishing low and high frequencies \cite{Inglis2016}. However, due to the inherent high dimensionality of the configuration space of fluid solvers these methods are still computationally challenging, making detailed fluid control hard to achieve. Additionally, they require the specification of target shapes for control, and automatic stylization of fluid features is not possible. Guided simulation control can also be used to reconstruct target smoke images. Okabe et al. \shortcite{okabe2015fluid} proposed an appearance transfer method for image-based \threeD~reconstruction of smoke volumes, while Eckert et al. \shortcite{Eckert2018} utilized proximal operators to reconstruct both the fluid density and motion from single or multiple views.




\vspace{1.5mm}
\emph{Machine Learning \& Fluids.} Combining fluid simulation with machine learning was first demonstrated by Ladick\'{y} et al. \shortcite{Ladicky2015}. The authors modeled a Lagrangian-based fluid solver by employing Regression forests to approximate particle positions and velocities given a neighborhood configuration. CNN-based architectures were employed in Eulerian contexts to substitute the pressure projection step \cite{Tompson2016, Yang2016} and to synthesize flow simulations from a set of reduced parameters \cite{Kim2018}. \revise{An} LSTM-based \cite{Wiewel2018} approach predicted changes on pressure fields for multiple subsequent time-steps. Closer to our work, Chu and Thuerey \shortcite{Chu2017} enhance simulations with patch correspondences between low and high resolution simulations. The patches are automatically created in a low resolution simulation, and then advected and deformed by the underlying flow field. A temporally coherent Generative Adversarial Network (GAN) was designed for smoke simulation super-resolution \cite{xie2018tempogan}, removing the reliance on Lagrangian tracking of features of the previous approach. Their method can produce detailed, high-quality results, but it does not support transfer of different smoke styles.

\vspace{1.5mm}
\emph{Neural Style Transfer} is the process of rendering image content in different styles by exploring CNNs. The seminal work of Gatys et al. \shortcite{gatys2015neural} was the first to transfer painting styles to natural images. Their model relies on extracting the content of an image by measuring filter responses of a pre-trained CNN, while modelling the style as summary feature statistics. The network's filter responses decompose the image complexity into multiple levels, ranging from low-level features to high-level semantics. Given a target style, NST approaches optimize CNN feature distributions of a source image style, while keeping its original content. Ruder et al. \shortcite{ruder2016artistic} implemented style transfer for video sequences, addressing temporal coherency issues due to occluded regions and long term correspondences, while Mordvintsev et al. \shortcite{mordvintsev2018differentiable} discuss the impact of different choices of image parameterizations for NST. For a detailed review of NST methods we refer to~\cite{jing1705neural}.



\vspace{1.5mm}
\emph{Differentiable rendering} allows the computation of derivatives of image pixels with respect to the variables used for rendering the image, \eg vertex positions, normals, colors, camera parameters, etc. These derivatives are crucial to optimization, inverse problems and deep learning backpropagation. Loper and Black \shortcite{Loper2014} proposed the first raster-based fully differentiable rendering engine with automatically computed derivatives. Anisotropic probing kernels were used to project \threeD~volumetric data similarly to x-ray scans \cite{qi2016volumetric}. Tulsiani et al. \shortcite{Tulsiani2017} used a differentiable ray consistency approach to leverage different types of multi-view observations which can vary from depth and color to foreground masks and normals. Differentiable volume sampling was implemented by Yan et al. \shortcite{Yan2016} to obtain \twoD~silhouettes from \threeD~volumes, adopting a similar sampling strategy as spatial transformer networks \cite{Jaderberg2015}. Kato et al. \shortcite{kato2018neural} and Liu et al. \shortcite{liu2018paparazzi} proposed a raster-based differential rendering for meshes with approximate and analytic derivatives, respectively. Recently, there is a growing interest on differentiable ray marching. Li et al. \shortcite{Li2018} introduces the first general-purpose differentiable ray tracer by removing discontinuities that appear when including visibility terms by directly sampling Dirac delta functions, while a differentiable path-tracer for inverse volumetric rendering with joint estimation of geometry was proposed by Velinov et al. \shortcite{Velinov2018}. Unlike previous approaches, our proposed differentiable renderer is the first to specifically tackle volumetric data stylization.

\section{Transport-based Neural Style Transfer}
\label{sec:Method}

Our method employs pre-trained CNNs for natural image classification as both feature extractor and synthesizer. As an alternative, we considered CNNs trained on synthetic \threeD~representations such as voxels~\cite{wu20153d}, meshes~\cite{masci2015geodesic} and point clouds~\cite{qi2017pointnet}. However, CNNs trained on \twoD~natural images have seen richer and denser information as there is a more expressive incidence of high-frequency features \cite{qi2016volumetric}, and data-sets have been thoroughly analyzed in terms of interpretability. Thus, as a classification CNN gets deeper, it shows its hierarchical interpretation of natural images organized from low-level patterns to high-level semantics~\cite{olah2017feature}.

The original neural style transfer (NST) \cite{gatys2015neural} transforms an initial noise image $I$ to match the content ($I_c$) and style ($I_s$) of input target images. The content loss $\mathcal{L}_c$ measures selected filter responses from a pre-trained classification CNN, while the style loss $\mathcal{L}_s$ measures the difference between specific filter's statistical distributions. The neural style transfer solves the optimization of
\begin{equation}\label{eq:objectiveGatys}
\begin{split}
\hat{I} & = \argmin_I \alpha \mathcal{L}_c(I, I_c) + \beta \mathcal{L}_s(I, I_s),\\
\end{split}
\end{equation}
where the weights $\alpha$ and $\beta$ control how the content and style modify the initial image $I$ along the optimization process.

\begin{figure}[t!]\centering
	\newdimen\gradFlowerHeight 
   \settoheight{\gradFlowerHeight}{\includegraphics[width=0.20\textwidth]{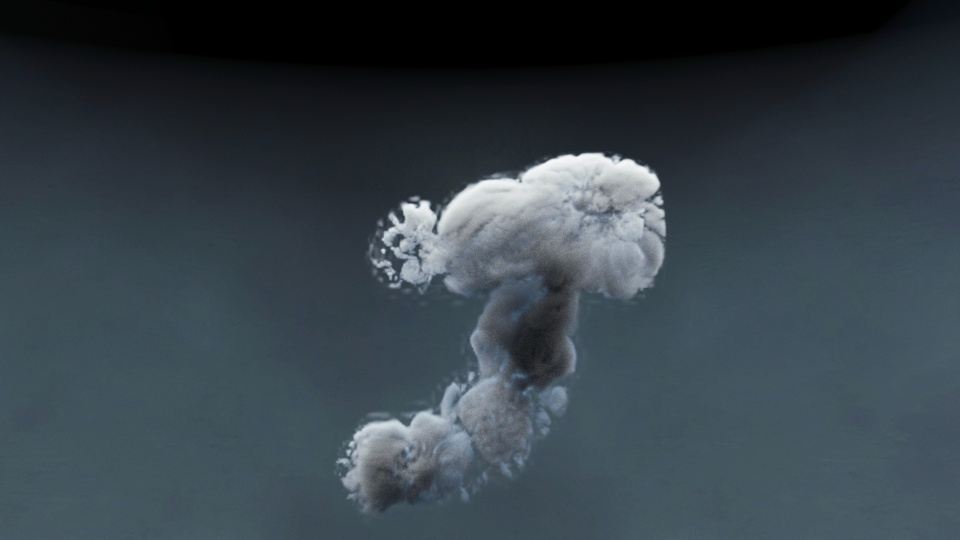}}
   \includegraphics[trim=310px 10px 270px 120px, clip, width=0.20\textwidth]{img/gradFlower}
   \includegraphics[trim=310px 10px 270px 120px, clip, width=0.20\textwidth]{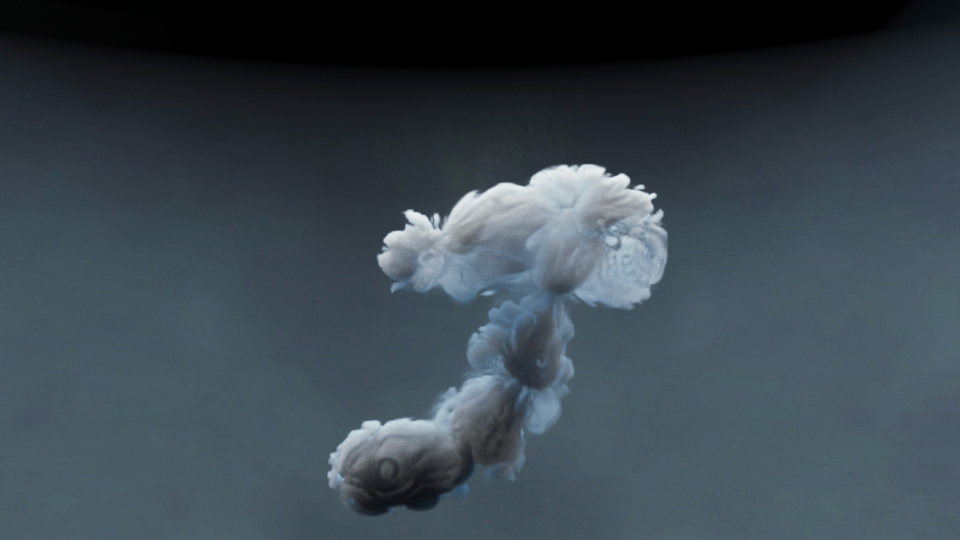}
   \hspace{-30px}
   \includegraphics[trim=0px 0px 0px 0px, clip, width=0.05\textwidth]{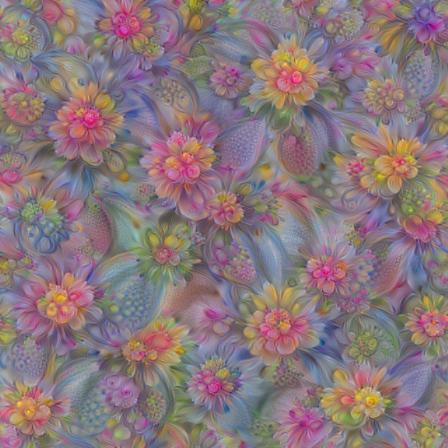} 

    \caption{Value-based (left) against transport-based density optimization (right) \revise{with a flower motif$^0$}. The value-based approach used in traditional image stylization approaches produces ghosting artifacts and thinner smoke structures, since density sources can be created and removed to match targeted features.}
	\label{fig:gradVsVelocityComparison}
    \vspace{-5px}
\end{figure}

Applying existing NST methods to stylize smoke data will lead to arbitrary creation of sources, since the volumetric density field is evaluated as an intensity image. Thus, we propose a transport-based neural style transfer (TNST), in which the stylization is driven by velocity fields instead of direct pixel / voxel corrections \revise{as introduced in \Fig{pipeline}}. The transport-based approach yields more degrees of freedom than directly changing the densities; specifically, it will yield a vector field while a standard value-based approach will output a scalar field correction. This is particularly useful in \threeD, since the directional information encoded by the vector field will be used to merge stylizations from distinct camera viewpoints. Additionally, this approach enables the control over smoke density sources and sinks during stylization: we implement a divergence control through the decomposition of the stylization vector field into its incompressible and irrotational parts. A comparison between value- and velocity-based stylizations is shown in \Fig{gradVsVelocityComparison}.


\begin{table}
\caption{Symbols, operators and configurable parameters}
\vspace{-8pt}
\begin{center}
 \begin{tabular}{c ? l}
 $\mathcal{L}_c, \mathcal{L}_s$ & \tablefont{Content and style losses} \\
 $\alpha, \beta$ & \tablefont{Weights controlling content and style losses} \\
 $d, \vec{u}$ & \tablefont{Input density and simulation velocity field} \\
 $d^*, \vec{v}$ & \tablefont{Stylized density and stylization velocity field} \\
 $\sigma$ & \tablefont{Density integrated spatially for a single frame} \\
 $\Phi, \vec{\Psi}$ & \tablefont{Irrotational and incompressible potentials} \\
 $\lambda$ & \tablefont{Weight between irrotational and incompressible vector fields} \\
 $\vec{p}_c, \vec{p}_s$ & \tablefont{Content and style input parameters} \\
 $\mathcal{R}, \mathcal{T}$ & \tablefont{Rendering and advection operators}\\
 $\mathcal{F}^l, \hat{\mathcal{F}}^l$ & \tablefont{CNN's spatial and flattened feature maps for layer $l$} \\
 $\mathcal{M}^l$ & \tablefont{User-defined feature map at layer $l$ for semantic transfer} \\
 $H, W, C$ & \tablefont{Height, width and channels of feature map or image} \\
 $\omega, w$ & \tablefont{Temporal coherency weight and window size} \\
 $\gamma$ & \tablefont{Transmittance absorption factor} \\
 $\vec{\theta}, \Theta$ & \tablefont{Individual viewpoint and set of viewpoints} \\
 $\eta$ & \tablefont{Learning rate size} \\
 \end{tabular}
\end{center}
\label{tab:param}
\vspace{-5pt}
\end{table}





\begin{figure}[t!]\centering
   \includegraphics[trim=310px 10px 270px 120px, clip, width=0.3\linewidth]{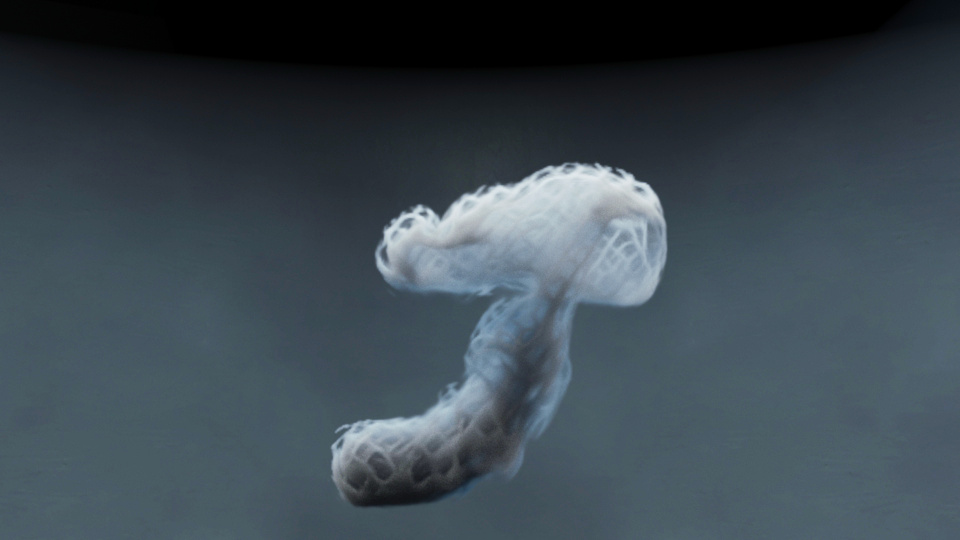}
   \includegraphics[trim=310px 10px 270px 120px, clip, width=0.3\linewidth]{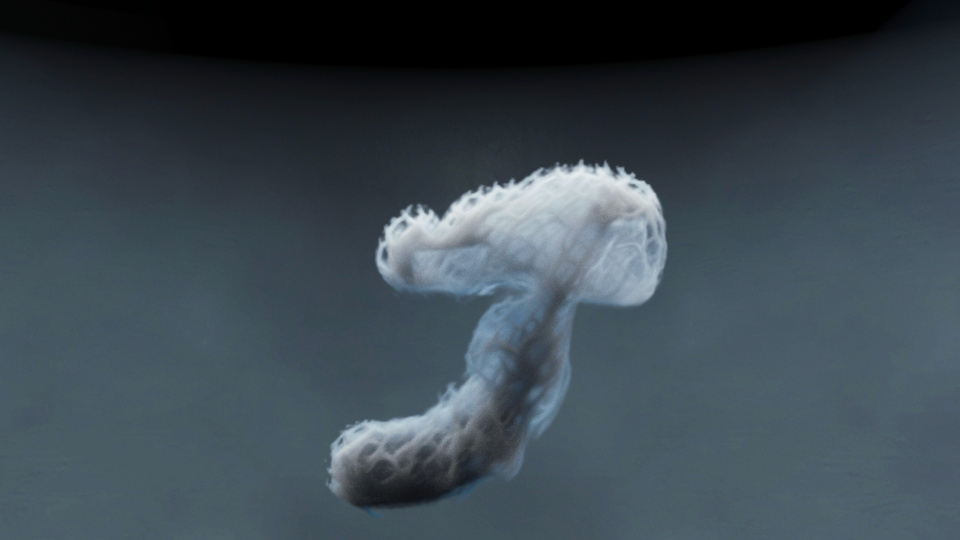}
   \includegraphics[trim=310px 10px 270px 120px, clip, width=0.3\linewidth]{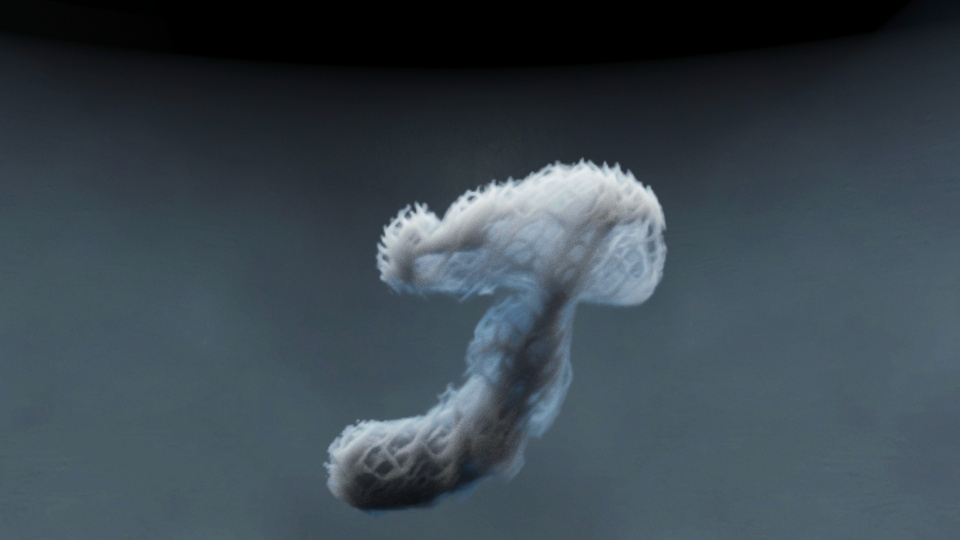}
   \hspace{-30px}
   \includegraphics[trim=0px 0px 0px 0px, clip, width=0.05\textwidth]{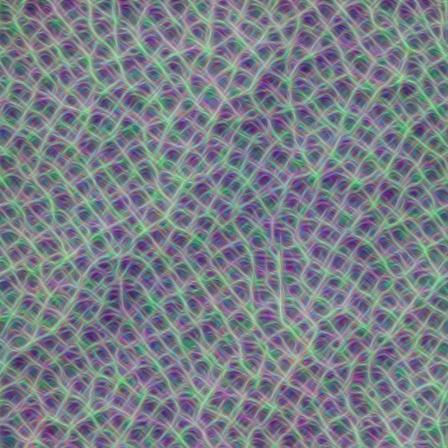}
    \caption{Results from semantic transfer of a net structure$^0$. Irrotational (left), mixed (middle) and incompressible (right) velocity fields.}
	\label{fig:fieldComparison}
    \vspace{-10px}
\end{figure}


\subsection{Single-frame Multi-view Stylization}
We define a single-frame loss for a given input image $I \in \varmathbb{R}^{H \times W}$
\begin{equation}\label{eq:loss}
\mathcal{L}(I, \vec{p}_c, \vec{p}_s) = \sigma \left [ \alpha \mathcal{L}_c(I, \vec{p}_c) + \beta \mathcal{L}_s(I, \vec{p}_s) \right ],
\end{equation}
where $\sigma = \sum_i^{H} \sum_j^{W} I_{ij}$ is the \revise{per-pixel} intensity $(I_{ij} \in [0, 1])$ integrated over the image, and $\vec{p}_c$ and $\vec{p}_s$ are user-specified parameters (Sections \ref{sec:contentLoss} and \ref{sec:styleLoss}) that control content and style transfers. We normalize the loss function by the integrated pixel intensities, since, contrary to natural images, pixels from our rendered depiction represent smoke intensities that will be used for stylization.



Given the input density field $d : \varmathbb{R}^3 \rightarrow \varmathbb{R}$ and a velocity field $\vec{v} : \varmathbb{R}^3 \rightarrow \varmathbb{R}^3$, the transport function $\mathcal{T}(d, \vec{v})$ advects $d$ by $\vec{v}$. Unlike image-based stylization, where pixels already contain color information of the represented image, our approach has to compute a valid rendering of the flow data. The renderer $\mathcal{R}_\theta(d)$ outputs a grayscale image (\Sec{renderer}) representing the density field for a specific viewpoint angle $\vec{\theta}$ from a discrete set of viewpoints $\Theta$. Our method optimizes a velocity field decomposed by a linear combination of its irrotational and incompressible parts by
\begin{equation}
\vec{v} = \lambda \nabla \Phi + (1 - \lambda) \nabla \times \vec{\Psi}
\label{eq:helmholtzVel}
\end{equation}
to achieve a desired stylized density field by minimizing
\begin{equation}\label{eq:realObjective}
\begin{split}
\hat{\Phi}, \hat{\vec{\Psi}} & = \argmin_\vec{\Phi, \vec{\Psi}} \sum_{\theta \in \Theta} \mathcal{L}(\mathcal{R}_\theta(d^*) \,, \vec{p}_c \,, \vec{p}_s),
\end{split}
\end{equation}
where $d^* = \mathcal{T}(d, \lambda \nabla \Phi + (1 - \lambda) \nabla \times \vec{\Psi})$ is the density field evolving towards stylization. Since our formulation optimizes for the scalar and vector potentials that transport the smoke, we allow the user to have direct control over the divergence of the stylization velocity field. Incompressible and irrotational velocity fields generate artistically different results and \Fig{fieldComparison} shows a comparison between both approaches. In order to create \threeD~structures, contributions from individual viewpoints $\mathcal{R}_\theta$ are summed, similarly to \cite{liu2018paparazzi}. We will discuss camera view sampling and renderer specifications in \Sec{renderer}. The next sections describe the loss functions that we use for semantic ($\mathcal{L}_c$) and style transfers ($\mathcal{L}s$).

\subsection{Semantic Transfer}
\label{sec:contentLoss}
Inspired by DeepDream\footnote{https://github.com/tensorflow/examples/tree/master/community/en/r1/deepdream.ipynb}, our method allows novel semantic transfer for stylizing smoke simulations by manipulating the content represented by the smoke. For example, smoke densities can be modified to portray patterns and shapes, such as squares or flowers, as depicted in \Fig{semanticTransfer}. Let $\mathcal{F}^l(I) \in \varmathbb{R}^{H_l \times W_l \times C_l}$ denote a feature map of $[H_l, W_l]$ dimensions with $C$ channels at the layer $l$ of the network with respect to an input image $I$.
The user-specified parameter $\vec{p_c}$ consists of an array of feature maps $\mathcal{M} \in \varmathbb{R}^{H_l \times W_l \times C_l}$ for all layers $l \in L$ specified by the array. We then define the content loss as to match features of a density field rendered image to a user-defined feature map by
\begin{equation}\label{eq:semanticLoss}
\mathcal{L}_c(I, \vec{p}_c) = \sum_l^{L} \left [ \frac{1}{H_l W_l C_l} \sum_i^{H_l} \sum_j^{W_l} \sum_k^{C_l} \left(\mathcal{F}^l_{ijk}(I) - \mathcal{M}^l_{ijk}\right)^2 \right ],
\end{equation}
where $\mathcal{F}^l_{ijk}(I)$ denotes an activated neuron of the CNN respective to the input image $I$ at position $(i,j)$ of the feature map's $k^\text{th}$ channel. The feature map $\mathcal{M}$ represents semantic features that will be transferred to the smoke (\eg flowers), and it controls the abstraction level of structures created in the stylization process. Choosing a feature map that lies on deeper levels of the network will create more intricate motifs, as shown in \Fig{semanticTransfer}. The user can choose the abstraction level for the semantic transfer to match the specific content of an input image by selecting shallow levels of the network layers; or, conversely, match classification textual tags, enabling a stylization that maximizes tags (\eg "volcano") on the output smoke.

Differently from previous image stylization approaches, we do not enforce the matching of content loss to the original unstyled image, and instead, the content loss is used to drive the flow data towards the creation of patterns. Since the smoke is modified by advecting its density towards stylization, we can guarantee that each iteration of the optimization will only slightly modify the original smoke by normalizing \revise{the gradients for updating} the velocities with the \revise{fixed} learning rate size ($\eta$).

\subsection{Style Transfer}
\label{sec:styleLoss}
\begin{figure}[t!]\centering
    \includegraphics[trim=330px 0px 330px 240px, clip, width=0.3\linewidth]{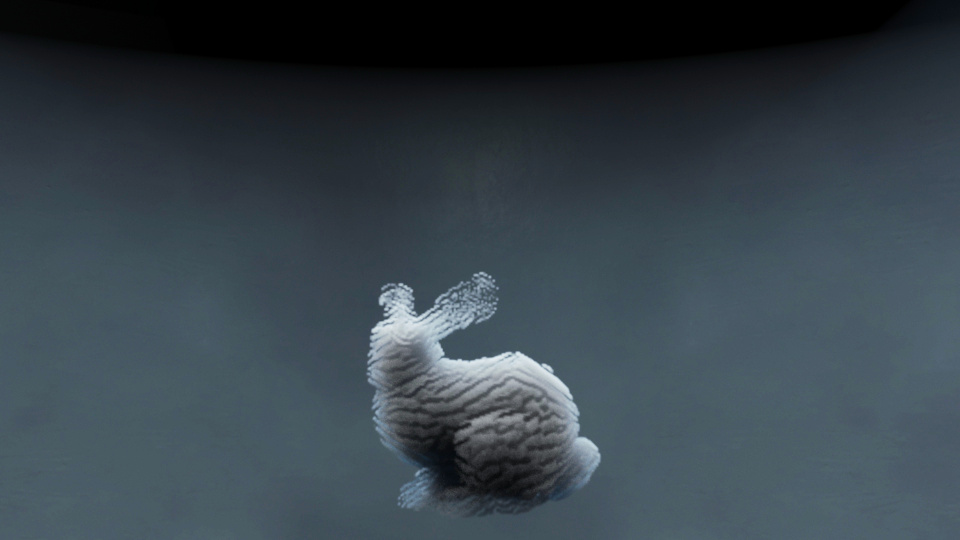}    
    \hspace{-30px}
    \includegraphics[trim=128px 128px 128px 128px, clip, width=0.05\textwidth]{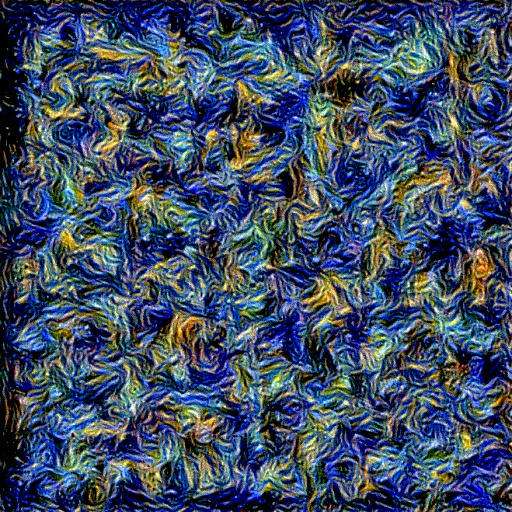}   
    \includegraphics[trim=330px 0px 330px 240px, clip, width=0.3\linewidth]{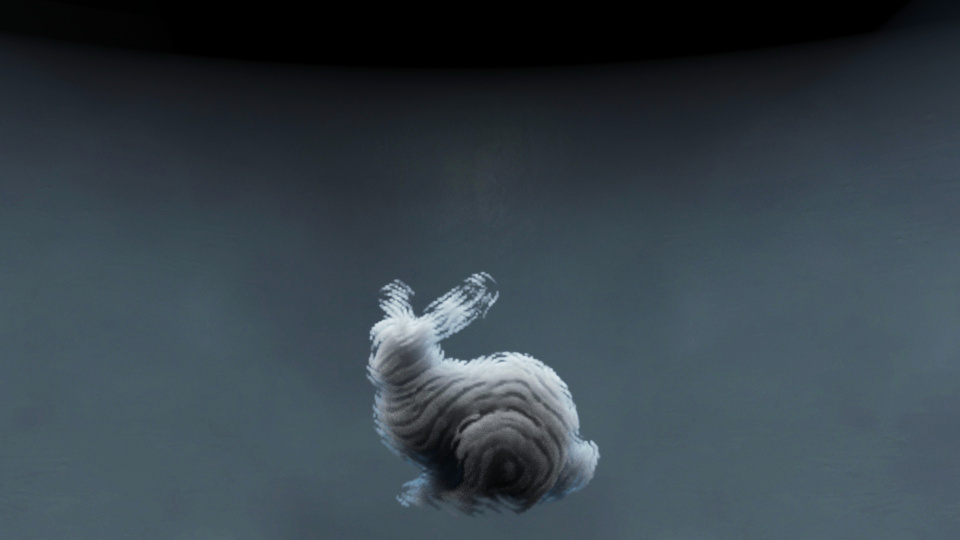}
    \hspace{-30px}
    \includegraphics[trim=128px 128px 128px 128px, clip, width=0.05\textwidth]{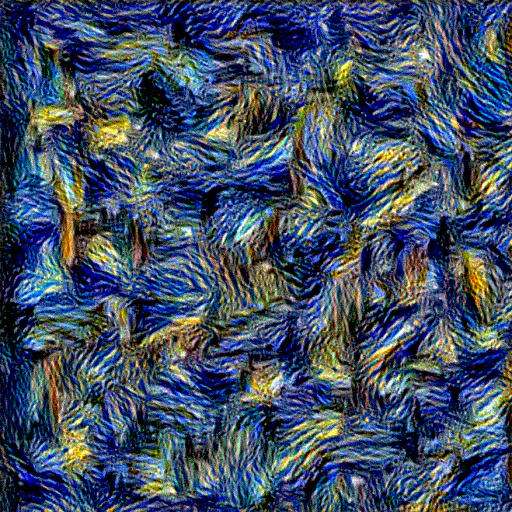}
    \includegraphics[trim=330px 0px 330px 240px, clip, width=0.3\linewidth]{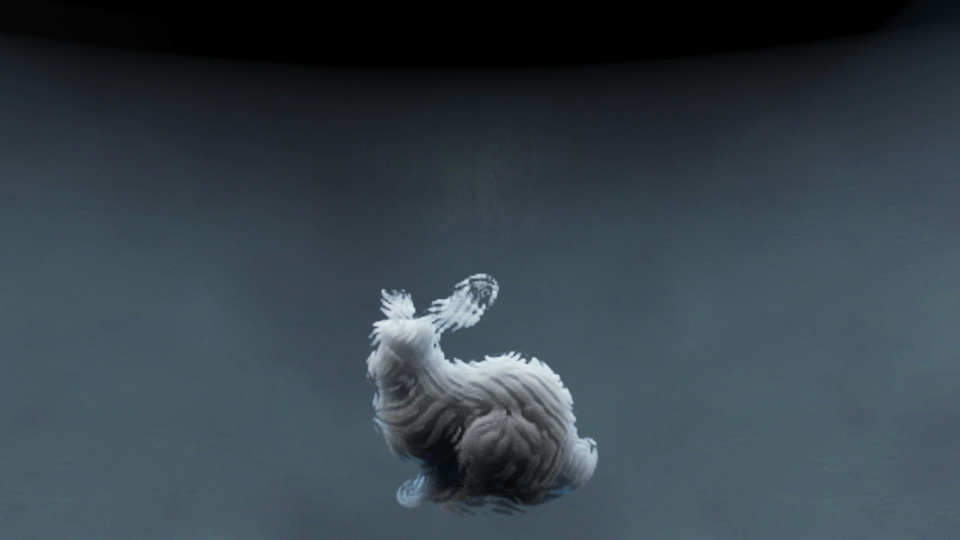}
    \hspace{-30px}
    \includegraphics[trim=128px 128px 128px 128px, clip, width=0.05\textwidth]{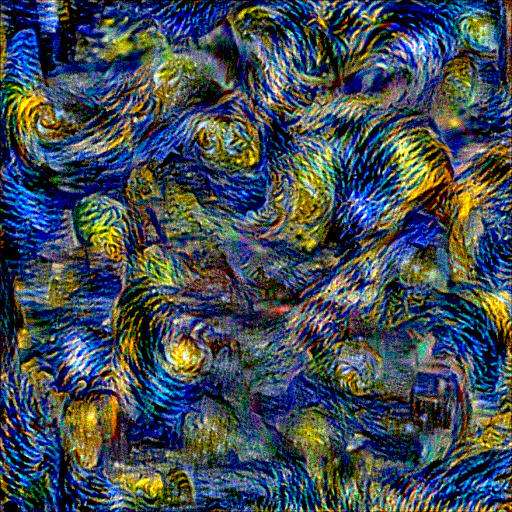}
    \caption{Abstraction levels of style features and their impact on the smoke stylization result. We can control low (left), medium (center) and high (right) levels of features. The \revise{corner images show} style representations corresponding to different feature levels.}
	\label{fig:SemanticAbstractionLevel}
    \vspace{-10pt}
\end{figure}
In addition to semantic transfer, \revise{which is designed to use "built-in" features of the pre-trained network \emph{without} any reference image as input,} our method allows the incorporation of a given input image style, as shown in \Fig{styleTransfer}. The style is computed by correlations between different filter responses, where the expectation is taken over the spatial extension of the input image. Hence, in contrast to semantic transfer, we minimize the difference between feature distributions. Given $\mathcal{\hat{F}}^l_k (I)$, which is the flattened one-dimensional version of a \twoD~filter map at the $k^\text{th}$ channel, the Gram matrix entry for two channels $m$ and $n$ is
\begin{equation}
G^l_{mn}(I) = \sum_{i}^{H_l\times W_l} \mathcal{\hat{F}}_{mi}^l(I) \; \mathcal{\hat{F}}_{ni}^l(I),
\end{equation}
where $i$ iterates over all pixels of the vectorized filter. Thus, the Gram $G^l(I)$ matrix of a $l^\text{th}$ layer has dimensions $C_l \times C_l$. The Gram matrix computes the dot product between all filter responses from a layer, storing correspondences of channels denoted by the row and column of an entry. The user-specified parameter $\vec{p_s}$ consists of a target image $I_s$ and a set of layers for which the style will be optimized for. Thus, the normalized loss function $\mathcal{L}_s$ for matching styles between an input image and a target style image is
\begin{equation}
\mathcal{L}_s(I, \vec{p}_s) = \sum_l^{L} \left [ \frac{1}{4C_l^2 (H_l\times W_l)^2} \sum_{m, n}^{C_l} \left( G^l_{mn}(I) - G^l_{mn}(I_s) \right)^2 \right ].
\label{eq:styleTransferFinal}
\end{equation}
%
Similarly to our semantic transfer, the Gram matrix layer choice in \Eq{styleTransferFinal} controls different abstraction levels of the stylization, as illustrated in Figure~\ref{fig:SemanticAbstractionLevel}. However, the style transfer does not match features that have spatial correlations relative to the input image, but rather \revise{approximates} filter response statistics. We further highlight differences between semantic and style transfer in Section~\ref{sec:semanticStyleTransfer}.



\subsection{Time-Coherent Stylization}
\begin{figure}[t!]
\centering
\includegraphics[trim=50px 60px 80px 20px, clip, width=0.47\textwidth]{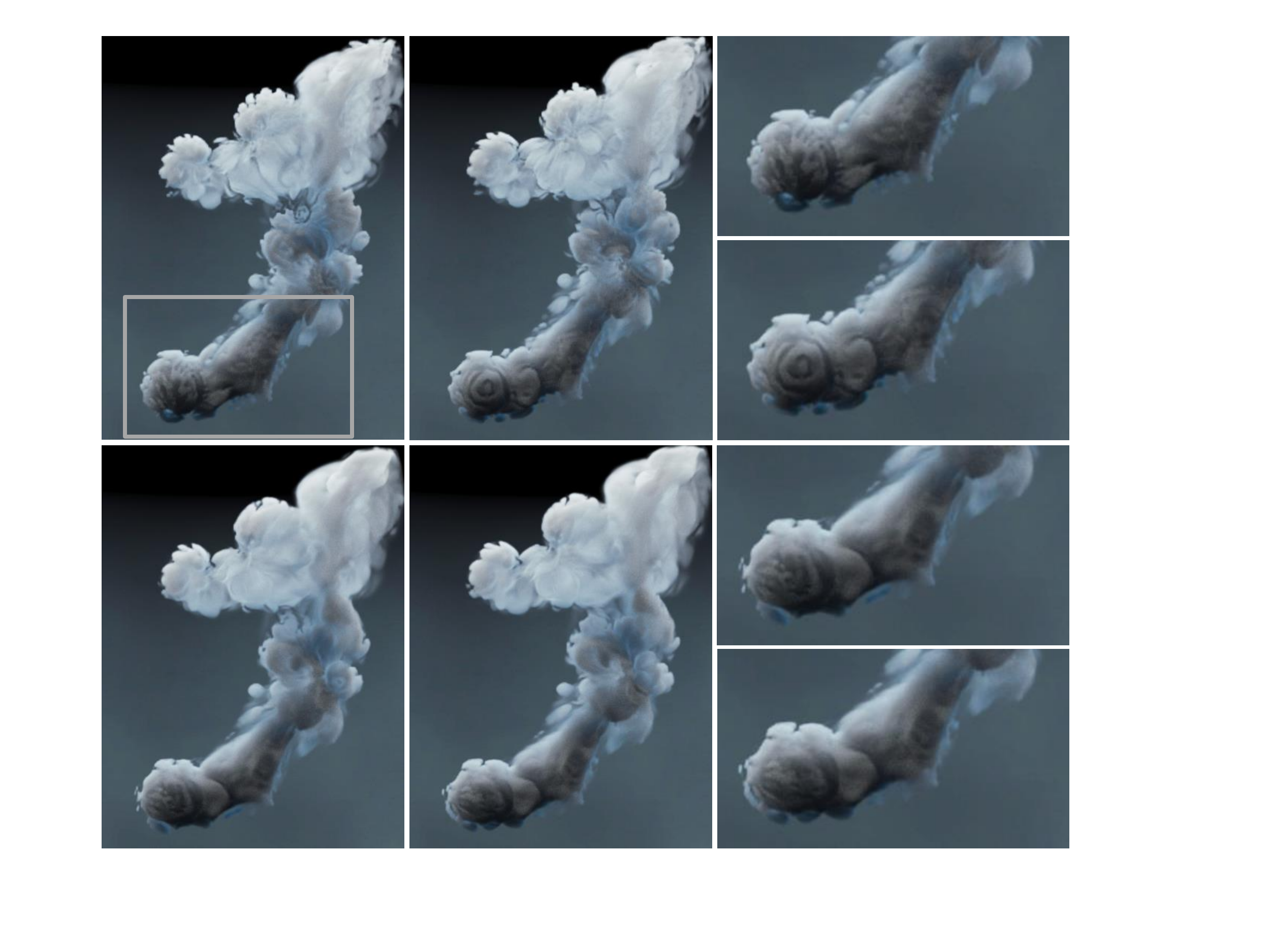}
             \caption{
             Two subsequent frames of the smoke jet example stylized with \revise{a flower$^0$ motif, }no time coherence (top) and window size 9 (bottom). For each frame, a close-up view corresponding to the highlighted region is shown on the right.
         Using our algorithm with a bigger window size ensures that the structures created in one frame are propagated to subsequent stylizations.}
    \label{fig:temporalCoherenceWindowSize}
    \vspace{-10pt}
\end{figure}
As densities are updated with the simulation advancement, distinct features can be emphasized by semantic and style transfer losses over different frames. Thus, flickering will occur if time-coherency between frames is not enforced explicitly, as shown in \Fig{temporalCoherenceWindowSize}. Given that the velocities of the original simulation transport densities over time, we use them to align stylization velocities computed independently for different frames. Once these velocities are aligned, we update a single frame stylization velocity field by smoothing subsequent aligned velocities together. Specifically, we define $\mat{U} = \{\vec{u}_{0}, \vec{u}_{1}, \ldots, \vec{u}_{n-1}, \vec{u}_{n}\}$ as the set of simulation velocities computed for the whole simulation duration. The advection function $\mathcal{T}_i^j$ that takes a stylization velocity at the $i^\text{th}$ frame to the $j^\text{th}$ frame is
\begin{equation} \label{eq:transportCases}
    \mathcal{T}_i^j(\vec{v}_{i}, \mat{U}) =
    \begin{cases}
        \mathcal{T}(\ldots \mathcal{T}(\mathcal{T}(\vec{v}_i,\vec{u}_i),\vec{u}_{i+1}) \ldots, \vec{u}_{j-1}),& \text{if } i < j\\
        \mathcal{T}(\ldots \mathcal{T}(\mathcal{T}(\vec{v}_i,-\vec{u}_{i-1}),-\vec{u}_{i-2}) \ldots, -\vec{u}_{j}),& \text{if } i > j\\
        \vec{v}_i,              & \text{if i = j}
    \end{cases},
\end{equation}
where $\mathcal{T}$ is a function that advects a velocity or a density field for a single time-step. \Eq{transportCases} is recursive, and aligning a velocity field defined $n$ frames away from a specific frame requires $n$ evaluations of the advection function. A temporally coherent velocity for stylization of frame $t$ is given as a linear combination of aligned neighbor velocity fields
\begin{equation}\label{eq:transport}
\vec{v}_t^\ast = \sum_{i=t-w}^{t+w} \omega_i \mathcal{T}_{i}^{t}(\vec{v}_i, \mat{U}),
\end{equation}
where $w$ is the number of neighboring frames evaluated in time, $2w + 1$ is the window size and $\omega_i$ is a weighting term. Let $\mat{V}_t = \{\vec{v}_{t - w}, \vec{v}_{t - (w - 1)}, \ldots, \vec{v}_{t}, \ldots, \vec{v}_{t + (w - 1)}, \vec{v}_{t + w}\}$ be the window of stylization velocities at time $t$ obtained by the combination of corresponding potential windows $\Phi_t, \vec{\Psi}_t$ defined in a range from $t-w$ to $t+w$. The time-coherent multi-view stylization optimization is
\begin{equation}\label{eq:objfinal}
\hat{\Phi}_t, \hat{\vec{\Psi}}_t = \argmin_{\Phi_t, \vec{\Psi}_t} \sum_{i=t-w}^{t+w} \sum_{\theta \in \Theta} \mathcal{L}(\mathcal{R}_\theta(\mathcal{T}(d_i,\vec{v}^\ast_i))\,, \vec{p}_c \,, \vec{p}_s).
\end{equation}

In practice, evaluating directly \Eq{objfinal} becomes \revise{infeasible} as the number of neighbors increases. The memory used by the automatic differentiation procedure to compute derivatives quickly grows as the window size increases. Thus, we approximate the solution of \Eq{objfinal} by first evaluating \Eq{realObjective} to find a set of stylization velocities computed for a single frame. Then, we merge the velocities per-frame individually using \Eq{transport}. This is performed iteratively for all simulation frames of a sequence, and the multi-view time-coherent process is summarized in \Algo{timeCoherence}.

\begin{algorithm}
\caption{Multi-view Time-coherent Smoke Stylization}
\begin{algorithmic}
\While{\small{$i < n_{iter}$}}
  \While{\small{$t < n_{frames}$}}
    \State{{\small{Compute density by $d^*_t = \mathcal{T}(d_t, \vec{v}_t^\ast $)}}}
    \For{\small{$\theta \in \Theta $}}
      \State{{\small{Render $I_t^\theta = \mathcal{R}_\theta(d_t^*)$ with angle $\theta$}}}
      \State{{\small{Obtain $\nabla \Phi^\theta_t, \nabla \vec{\Psi}^\theta_t$ from $\mathcal{L}(I_t^\theta, \vec{p}_c \,, \vec{p}_s)$}}}  
    \EndFor
    \State{{\small{Merge gradients from views $\nabla \Phi^\theta_t, \nabla \vec{\Psi}^\theta_t$ to obtain $\nabla \Phi^*_t, \nabla \vec{\Psi}^*_t$}}}
    \For{\small{$t_w = t - w, t_w < t + w$}}
        \State{{\small{Align gradients $\nabla \Phi^*_t, \nabla \vec{\Psi}^*_t$ to obtain $\nabla \Phi_t, \nabla \vec{\Psi}_t$ Eq.~\ref{eq:transportCases}}}}
    \EndFor
    \State{{\small{$\Phi_t = \Phi_t + \eta \nabla \Phi_t, \vec{\Psi}_t = \vec{\Psi}_t + \eta \nabla \vec{\Psi}_t$}}}
    \State{{\small{$\vec{v}_t^* = \lambda \nabla \Phi_t + (1 - \lambda) \nabla \times \vec{\Psi_t} $}}}
  \EndWhile
\EndWhile
\end{algorithmic}
\label{algo:timeCoherence}
\end{algorithm} 
\section{Differentiable Smoke Renderer}
\label{sec:renderer}
Similar to the flat shading approach proposed by Liu et al. \shortcite{liu2018paparazzi} for stylizing meshes, our smoke renderer is lightweight. The optimization of \Eq{realObjective} heavily relies on rendered density representations, and an overly sophisticated volumetric renderer compromises efficiency. Our renderer outputs grayscale images, in which pixel intensity values will correspond to density occupancy data. Thus, modelling smoke self-shadowing would map shadowed regions to empty voxels on the rendered image. Nevertheless, our results show that meaningful correspondences between the stylization velocities and density fields can be computed on representations that do not match perfectly the ones produced by the final rendered image.



\begin{figure}[t!]
\centering
\includegraphics[trim=15px 10px 0px 60px, clip, width=0.208\linewidth]{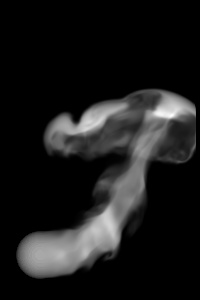}
\includegraphics[trim=310px 10px 270px 120px, clip, width=0.24\linewidth]{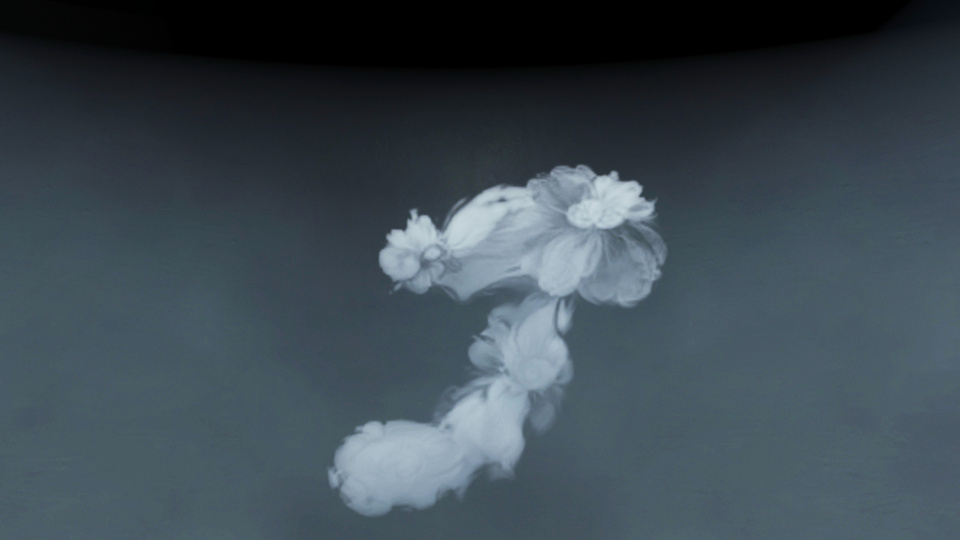}
\includegraphics[trim=310px 10px 270px 120px, clip, width=0.24\linewidth]{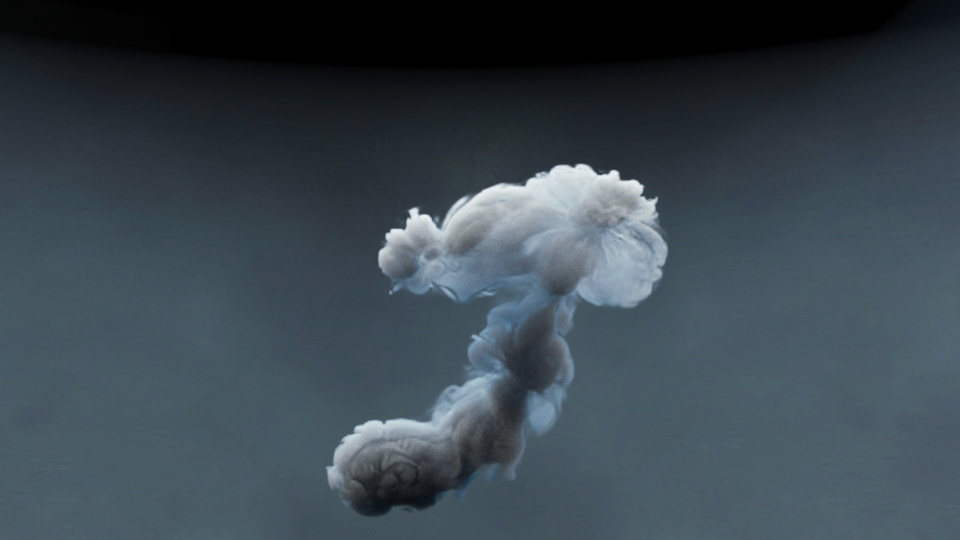}
\includegraphics[trim=310px 10px 270px 120px, clip, width=0.24\linewidth]{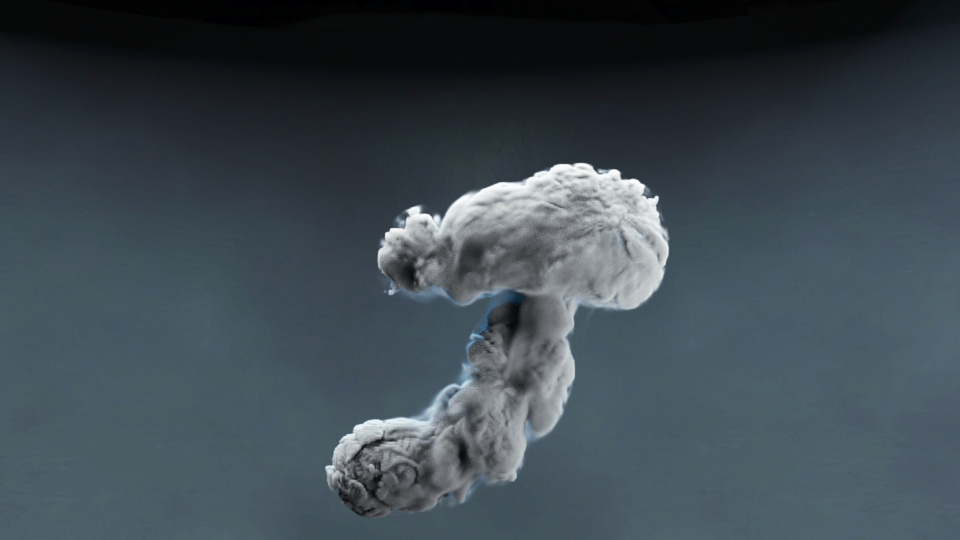}\\
\begin{subfigure}{0.208\linewidth}
\includegraphics[trim=15px 10px 0px 60px, clip, width=1\linewidth]{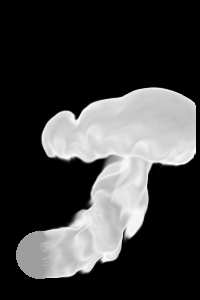}
\vspace*{-5mm}
\caption*{\scriptsize{Our Renderer}}
\end{subfigure}
\begin{subfigure}{0.24\linewidth}
\includegraphics[trim=310px 10px 270px 120px, clip, width=1\linewidth]{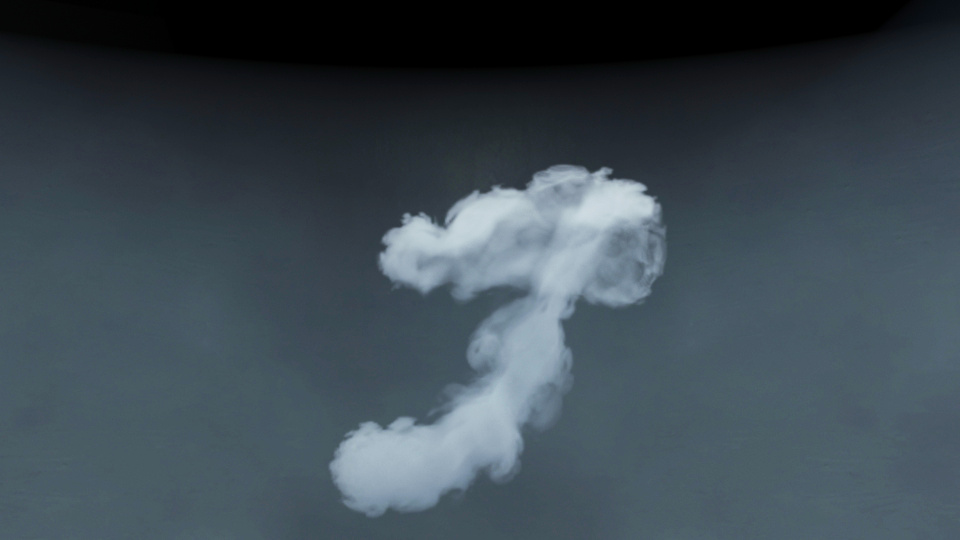}
\vspace*{-5mm}
\caption*{\scriptsize{Shallow Appearance}}
\end{subfigure}
\begin{subfigure}{0.24\linewidth}
\includegraphics[trim=310px 10px 270px 120px, clip, width=1\linewidth]{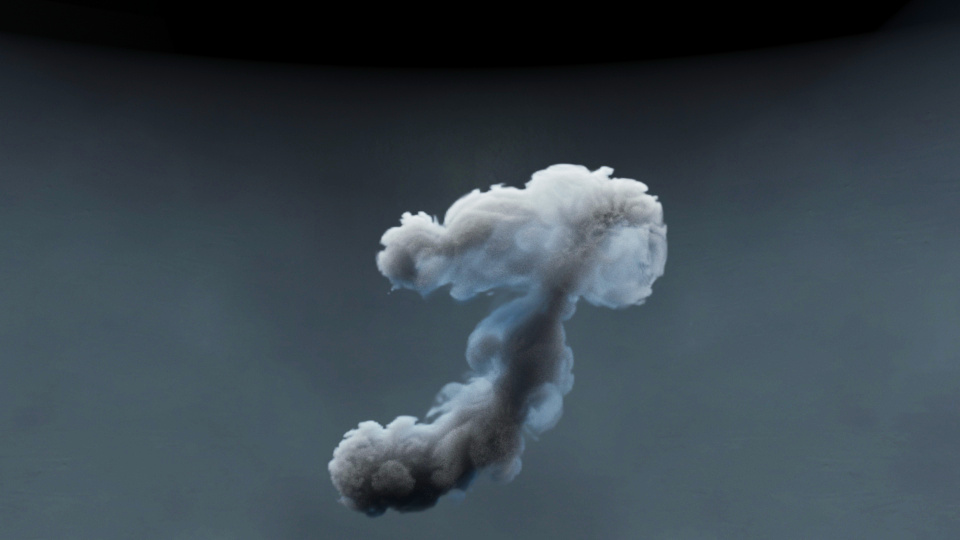}
\vspace*{-5mm}
\caption*{\scriptsize{Medium Appearance}}
\end{subfigure}
\begin{subfigure}{0.24\linewidth}
\includegraphics[trim=310px 10px 270px 120px, clip, width=1\linewidth]{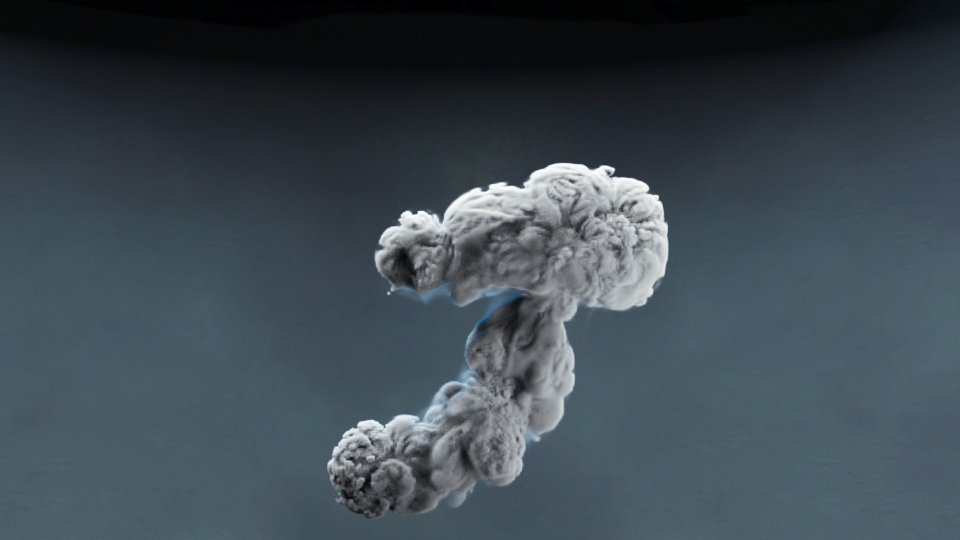}
\vspace*{-5mm}
\caption*{\scriptsize{Thick Appearance}}
\end{subfigure}
\caption{The value of $\gamma$ controls how the smoke density is stylized. Smoke images (left) produced by our renderer with $\gamma=0.01$ (top) and $\gamma=1$ (bottom). The final smoke renderer is configured with varying thickness, highlighting how the stylization gets transferred for different smoke appearances.}
\label{fig:renderDiffReflectance}
\vspace{-10pt}
\end{figure}


The smoke stylization optimization usually performs many iterations, computing derivatives of the loss function (\Eq{realObjective}) with respect to the velocity field by automatic differentiation. Therefore, the volumetric rendering requires efficiency. Our lightweight differentiable rendering algorithm only incorporates a single directional light traced directly from the pixel rendered from an orthographic camera. We measure how much of this single light ray gets transmitted through the inhomogeneous participating media, which is described by \cite{Fong2017}, to compute the transmittance and the image pixel grayscale value as
\begin{equation}
\begin{split}
\tau(\vec{x}, \vec{r}) = e^{-\gamma \int_{\vec{x}}^{\vec{r}_{max}} d(\vec{r}) \; dr} \\
I_{ij} = \int_{0}^{\vec{r}_{max}} d(\vec{x}) \, \tau(\vec{x}, \vec{r})\; dx ,
\label{eq:renderingEquation}
\end{split}
\end{equation}
where $\vec{r}_{ij}$ is a vector traced from pixel $ij$ into the normal direction of an orthographic camera, $d(\vec{\vec{r}}$) is the density value, $\gamma$ is a transmittance absorption factor, and $r_{max}$ is the maximum length of the traced ray.  The value computed at each image pixel is the integral of the transmittance multiplied by the density values, mapping empty and full smoke voxels to 0 and 1, respectively. We additionally multiply the transmittance and densities along the integration ray since it generates richer features for thicker smoke scenarios. Comparisons between this approach against simply integrating the transmittance along the view-ray are shown in our supplementary material.


The smoke density is linearly mapped to extinction using the scaling factor $\gamma$, which determines how quickly light gets absorbed by the smoke. \Fig{renderDiffReflectance} shows that minimizing the discrepancy between the final rendered smoke and the representation in which it is optimized is important. Setting low transmittance constants in the stylization renderer will result in more aggressive smoke modifications towards the normal view direction, while high transmittance will over-constrain the stylization velocity field to the smoke surface that is closer to the camera. \Fig{renderDiffReflectance} shows the effect of varying $\gamma$ values with the final rendered smoke thickness: low $\gamma$ values will produce better results for shallow smoke appearances, while higher $\gamma$ values will more efficiently stylize thicker smoke.

\subsection{Camera Design Specifications}

\begin{figure}[h!]
    \includegraphics[trim=75px 90px 75px 75px, clip, height=0.37\linewidth]{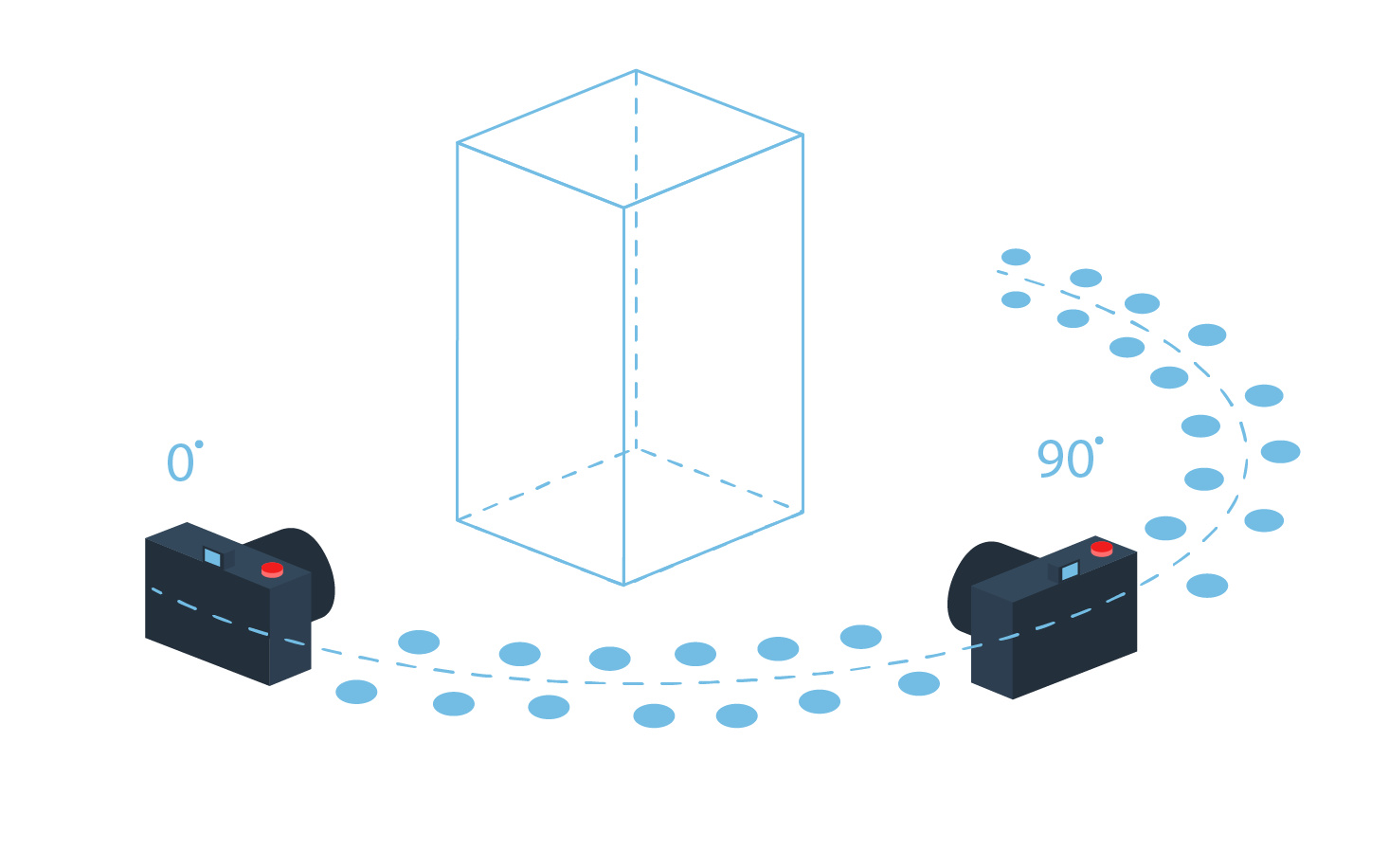}
    \hspace{10pt}
    \includegraphics[trim=500px 90px 500px 75px, clip, height=0.37\linewidth]{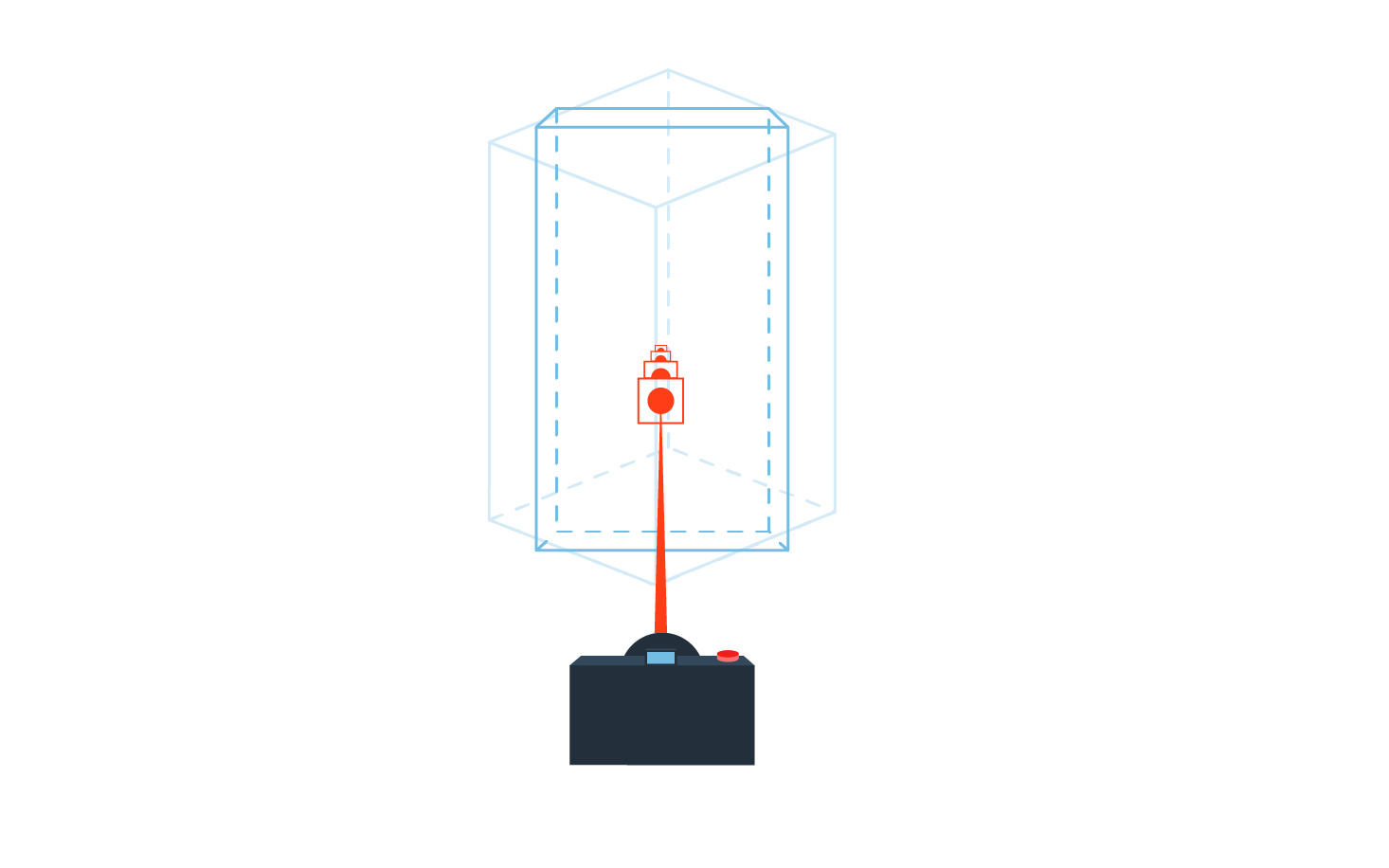}
    \caption{Multiview camera configuration. We sample a camera path with Poisson sampling, which prevents smoothing of density details between predefined viewpoints (left). The volumetric smoke grid is aligned with the camera viewpoint to facilitate light ray integration (right).}
	\label{fig:smokeMultiView}
    \vspace{-10pt}
\end{figure}

Participating media naturally incorporates transparency, and a single-view stylization update will be propagated inside the volumetric smoke even though the rendered image is two dimensional. Therefore it is not necessary to uniformly cover every viewpoint of the smoke with equal probability as in \cite{liu2018paparazzi}. Given a predefined camera path, we use Poisson sampling around a small area of its trajectory (\Fig{smokeMultiView}, left) to avoid bias that would be introduced by a fixed set of viewpoints.

Since feature maps obtained from \twoD~views of the camera are used, we specify that the image rendered by the camera is invariant to zooming, panning and rolling. This means that if the camera is moving (as in \Fig{smokeMultiView}, left), our renderer automatically centers the smoke representation in the frame, only responding to variations of the viewing angle. These invariances ensure that filter map activations remain constant as long as no new voxels are shown in the rendered image; rotations, however, have to be accounted for. Thus, our renderer camera position is parameterized by the polar coordinates tuple $\vec{\theta} = (\theta_1, \theta_2)$, while the camera always points to a fixed point inside the smoke. Note that this simplification is only possible since we are adopting an orthogonal camera, and a perspective projection might reveal new voxels with translational movement.
\setlength{\columnsep}{0pt}%
\begin{wrapfigure}[6]{hR}{0.55\linewidth} 
\vspace{-5mm}
\begin{center}
\begin{subfigure}{0.95\linewidth}
\centering
\includegraphics[trim=20px 15px 30px 135px, clip, width=0.48\linewidth]{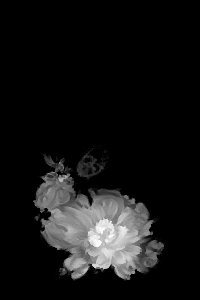}
\includegraphics[trim=20px 15px 30px 135px, clip, width=0.48\linewidth]{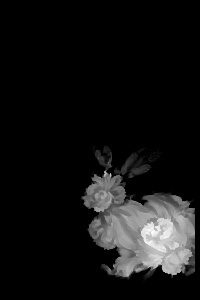}
\end{subfigure}
\end{center}
\label{fig:insetGradComparision}
\end{wrapfigure}
The inset image shows how enhanced features (\eg patterns at the bunny face) vary due to translations in the image space, in which the stylization should remain constant.

In order to evaluate \Eq{renderingEquation} for multiple perspectives, we need to integrate smoke voxels along the camera view direction. Implementing a classic ray-marching sampling along an arbitrary ray direction is challenging in Deep Learning frameworks, which are usually optimized for tensor operations. Thus, we adopted the spatial transformer network (STN) of Jaderberg et al. \shortcite{Jaderberg2015}. The STN instances a rotated \threeD~domain with the same dimensionality as the original one that is aligned with the camera view as illustrated in \Fig{smokeMultiView} (right). This allows us to evaluate samples by evoking simple built-in features that implement voxel summations along the view direction of each pixels' ray. These specifications make our rendering algorithm efficient, accounting for about $30\%$ of time taken for processing a batch (see \Tab{stat}).

\section{Results}
We demonstrate that our approach can reliably transfer various styles from images onto volumetric flow data, with automatic semantic instantiation of features and artistic style transfer. All our stylization examples employed a mask with soft edges that is extracted from the original smoke data. The mask is applied to the potential field, and it restricts modifications to be close to the original smoke border while enhancing temporal accuracy near smoke boundaries (\Sec{discussion}). We provide further results, extended masking discussion, influence of parameters and additional 2-D examples in our supplemental material. The advection operator $\mathcal{T}$  is implemented by the MacCormack method \cite{Selle2008}. We refer the reader to the supplemental video for the corresponding animations.

Equations~\ref{eq:semanticLoss} and~\ref{eq:styleTransferFinal} are optimized by stochastic gradient descent, with the gradients computed by backpropagation on GoogleNet \cite{Szegedy2014}. Although we use automatic differentiation, analytic differentiation would allow us to fit even bigger simulation examples \cite{liu2018paparazzi}. We modified the original stride size of GoogleNet's first layer from two to one\footnote{https://medium.com/mlreview/getting-inception-architectures-to-work-with-style-transfer-767d53475bf8} to remove checkerboard patterns that occur when the kernel size is not divisible by the CNN's stride size. We use a fixed learning rate and apply multiscale stylization and Laplacian pyramid gradient normalization techniques$^2$ for boosting lower frequencies. Parameters and performance values for all examples are summarized in \Tab{stat}. Our implementation uses \emph{tensorflow} evaluated on a TITAN Xp GPU (12GB). The input simulations have been computed with different solvers. We used \emph{mantaflow} for the smokejet and bunny examples in \Fig{semanticTransfer} and \Fig{styleTransfer}, Houdini for computing the volcano in \Fig{teaser}, and a dataset from Sato et al. \shortcite{sato2018example} in \Fig{satoTransfer}.


\subsection{Semantic and Style Transfers}
\label{sec:semanticStyleTransfer}
\begin{figure*}[t!]
 \centering
    \begin{subfigure}{0.15\textwidth}
    \caption*{\small{Original Simulation}}
    \end{subfigure}
    \begin{subfigure}{0.15\textwidth}
    \caption*{\imagefont{3b bottleneck, $c=44$}}
    \end{subfigure}
    \begin{subfigure}{0.15\textwidth}
    \caption*{\imagefont{3b bottleneck, $c=66$}}
    \end{subfigure}
    \begin{subfigure}{0.15\textwidth}
    \caption*{\imagefont{4b pool reduce, $c=16$}}
    \end{subfigure}
    \begin{subfigure}{0.15\textwidth}
    \caption*{\imagefont{4b pool reduce, $c=60$}}
    \end{subfigure}
    \begin{subfigure}{0.15\textwidth}
    \caption*{\imagefont{4b pool reduce, $c=38$}}
    \end{subfigure}
    \\
     \vspace{1pt}
    \includegraphics[trim=290px 0px 250px 110px, clip, width=0.15\textwidth]{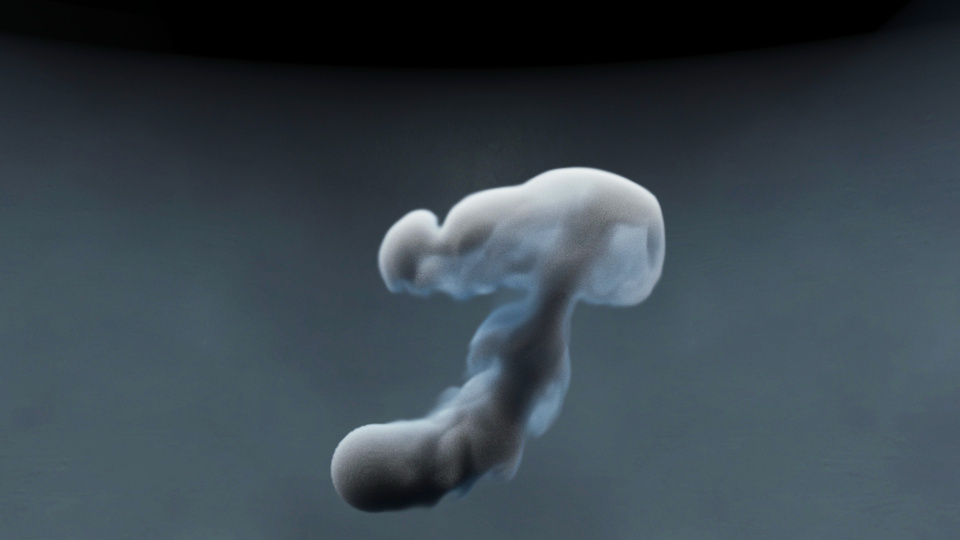}
    \includegraphics[trim=290px 0px 250px 110px, clip, width=0.15\textwidth]{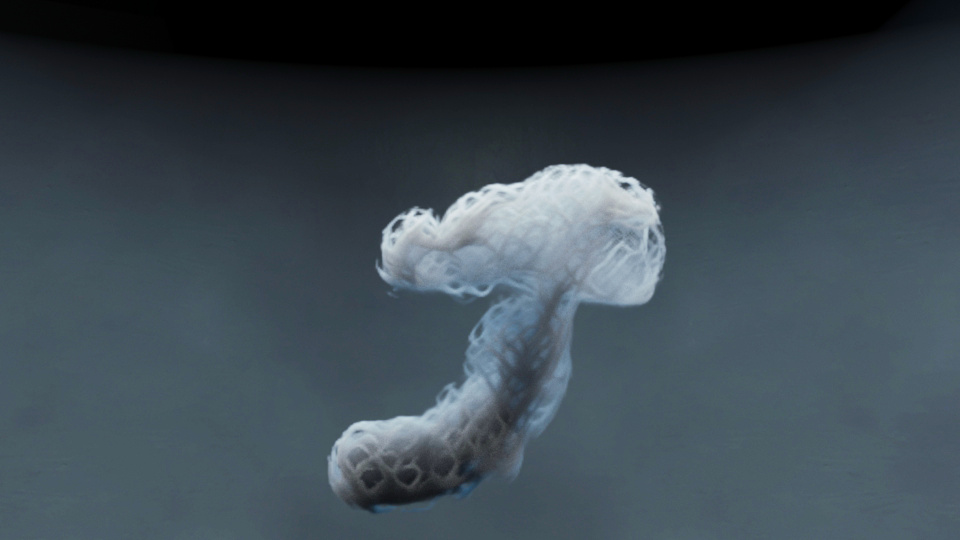}
	\hspace{-27.7px}\includegraphics[width=0.05\linewidth]{img/supp/patterns_mixed3b_3x3_bottleneck_pre_relu/44}
    \includegraphics[trim=290px 0px 250px 110px, clip, width=0.15\textwidth]{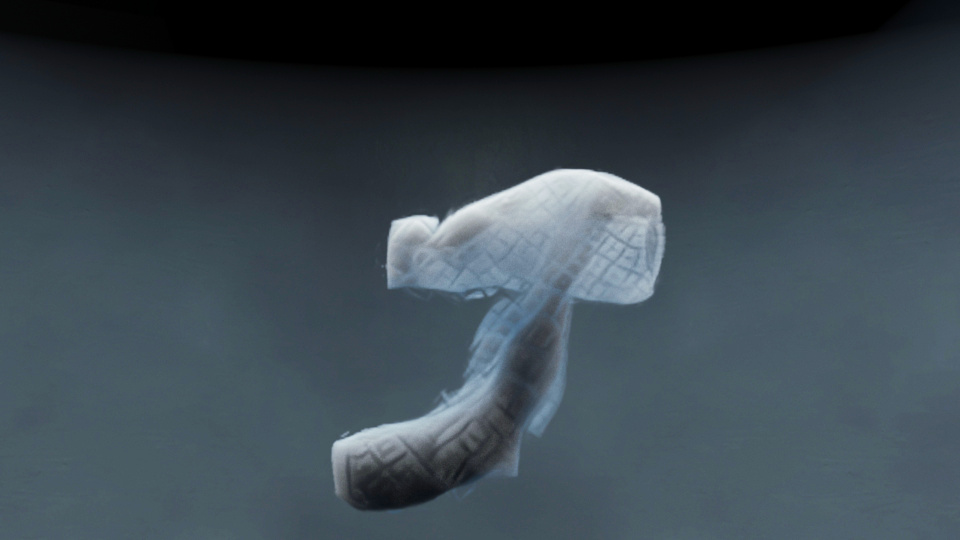}
    \hspace{-27.7px}\includegraphics[width=0.05\linewidth]{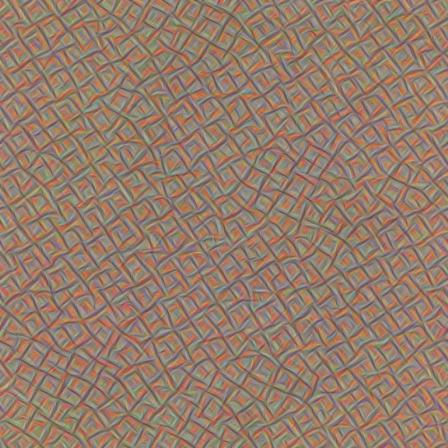}
    \includegraphics[trim=290px 0px 250px 110px, clip, width=0.15\textwidth]{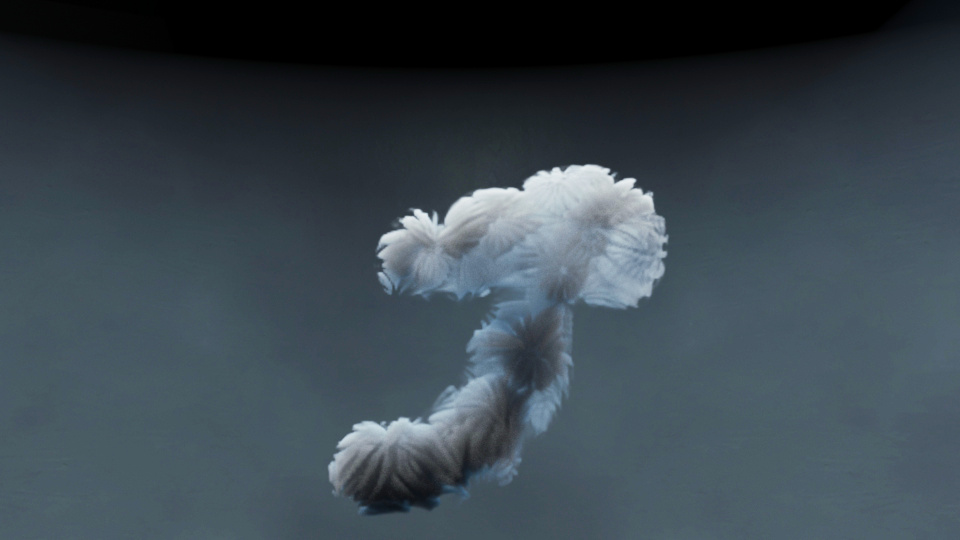}
    \hspace{-27.7px}\includegraphics[width=0.05\linewidth]{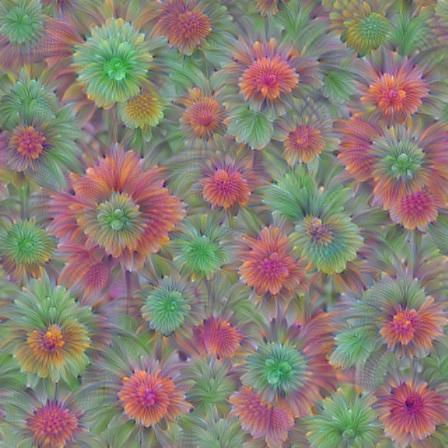}
    \includegraphics[trim=290px 0px 250px 110px, clip, width=0.15\textwidth]{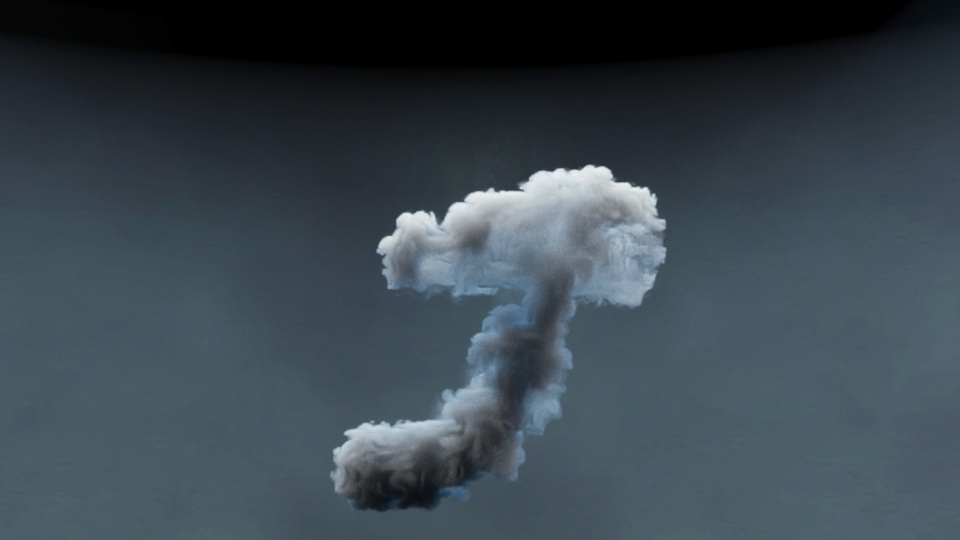}
    \hspace{-27.7px}\includegraphics[width=0.05\linewidth]{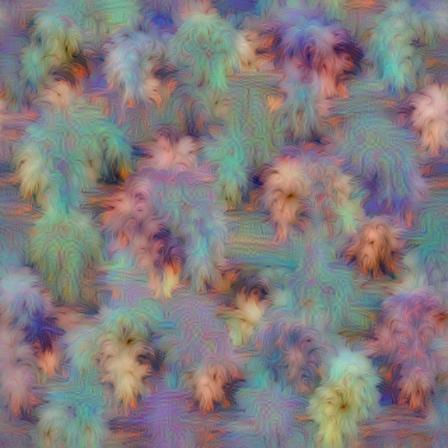}
    \includegraphics[trim=290px 0px 250px 110px, clip, width=0.15\textwidth]{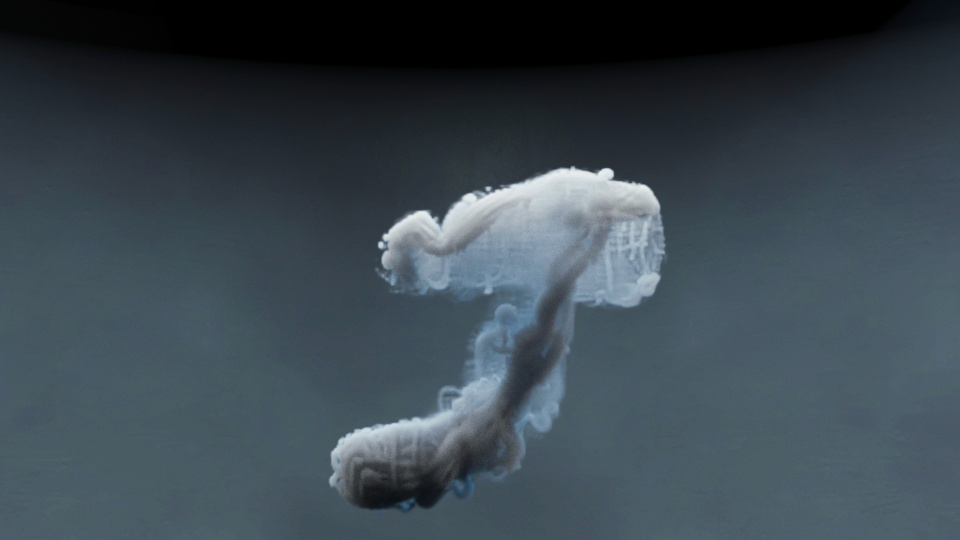}
    \hspace{-27.7px}\includegraphics[width=0.05\linewidth]{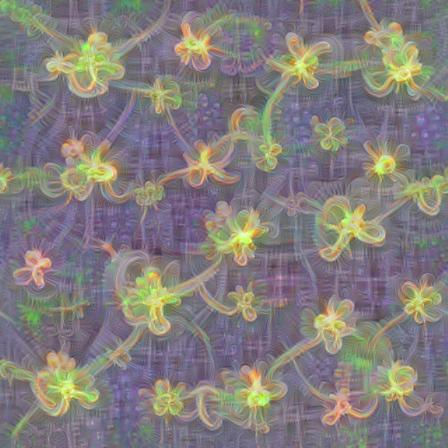}
    \\
     \vspace{1pt}
    \includegraphics[trim=320px 0px 320px 250px, clip, width=0.15\textwidth]{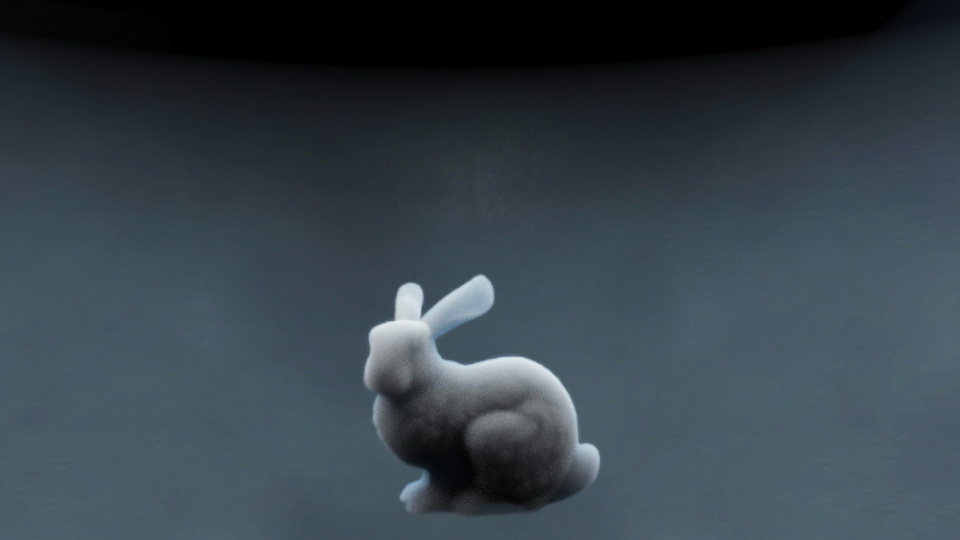}
    \includegraphics[trim=320px 0px 320px 250px, clip, width=0.15\textwidth]{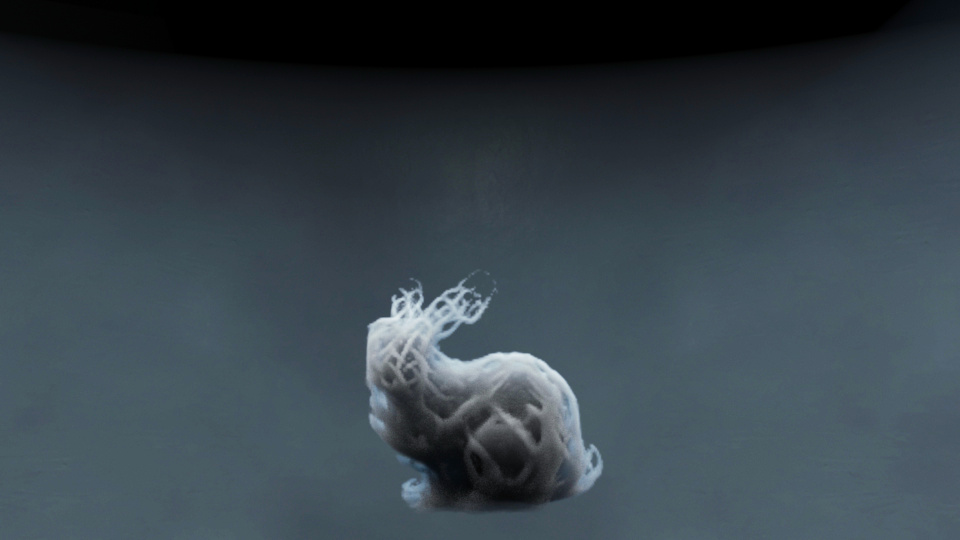}
    \includegraphics[trim=320px 0px 320px 250px, clip, width=0.15\textwidth]{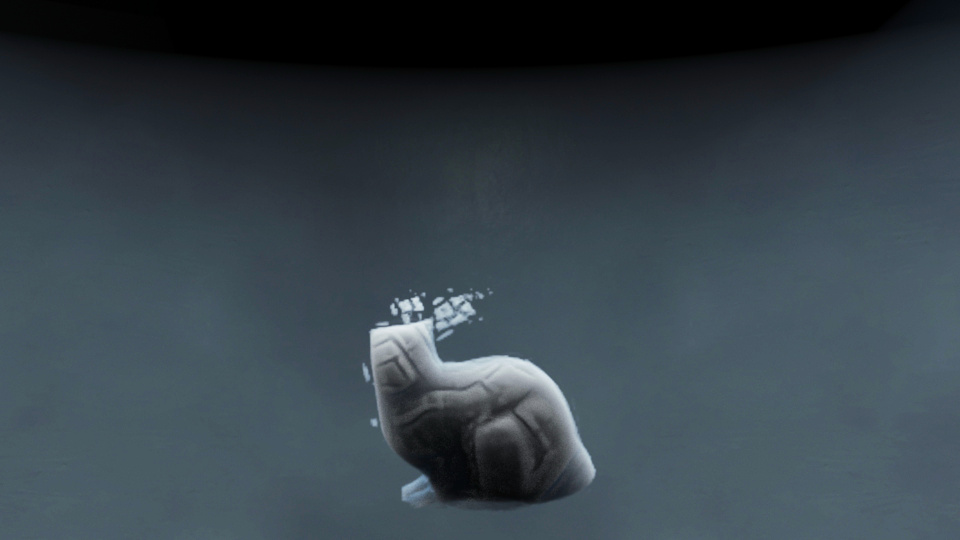}
    \includegraphics[trim=320px 0px 320px 250px, clip, width=0.15\textwidth]{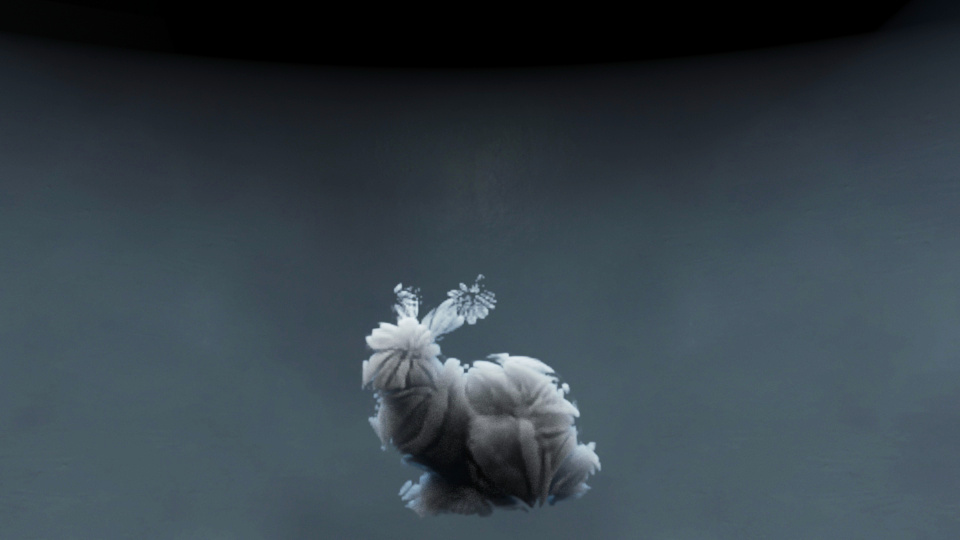}
    \includegraphics[trim=320px 0px 320px 250px, clip, width=0.15\textwidth]{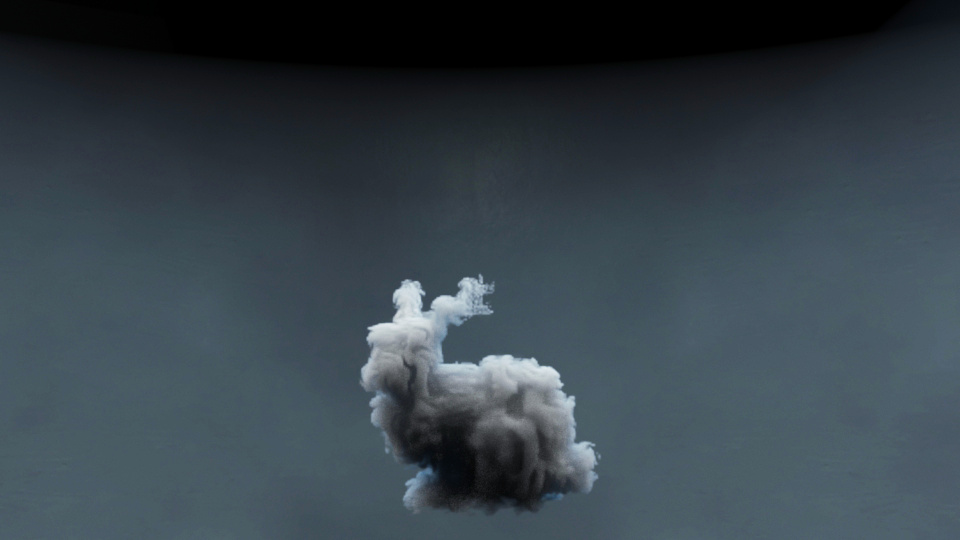}
        \includegraphics[trim=320px 0px 320px 250px, clip, width=0.15\textwidth]{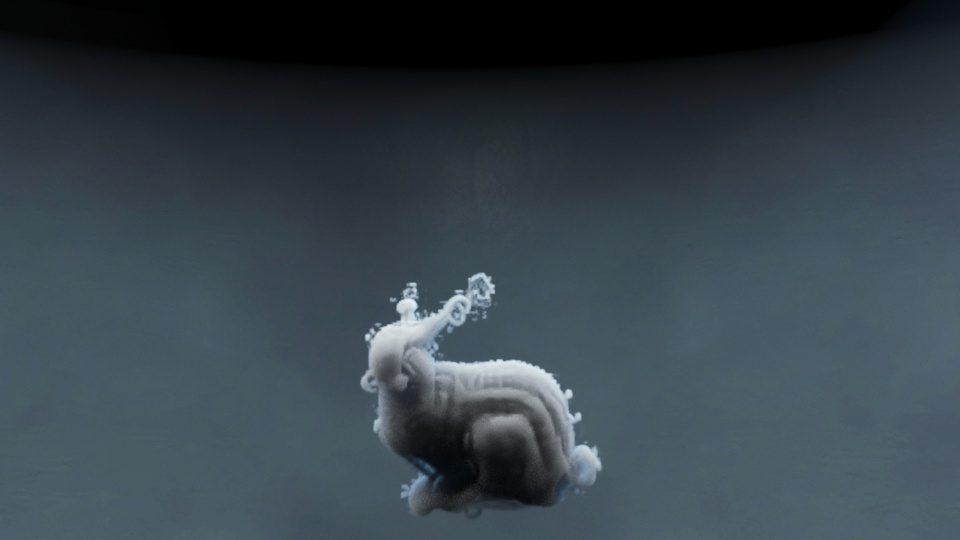}
    \\
    \caption{Semantic transfer applied to a smokejet and bunny simulations (leftmost column). Images on the same column are stylized with the feature map depicted on the \revise{right corner$^0$}. The examples for semantic transfer depict different levels of abstraction, showing patterns that occur at shallow levels of the network (first two columns) and intricate motifs that are represented at deeper levels (last three columns).}
\label{fig:semanticTransfer}
\end{figure*}

\begin{figure*}[t!]
 \centering
    \begin{subfigure}{0.15\textwidth}
    \caption*{\small{Fire}}
    \end{subfigure}
    \begin{subfigure}{0.15\textwidth}
    \caption*{\small{Volcano}}
    \end{subfigure}
    \begin{subfigure}{0.15\textwidth}
    \caption*{\small{Seated Nude}}
    \end{subfigure}
    \begin{subfigure}{0.15\textwidth}
    \caption*{\small{Starry Night}}
    \end{subfigure}
    \begin{subfigure}{0.15\textwidth}
    \caption*{\small{Blue Strokes}}
    \end{subfigure}
    \begin{subfigure}{0.15\textwidth}
    \caption*{\small{Spiral Patterns}}
    \end{subfigure}
    \\
     \vspace{1pt}
    \includegraphics[trim=290px 0px 250px 110px, clip, width=0.15\textwidth]{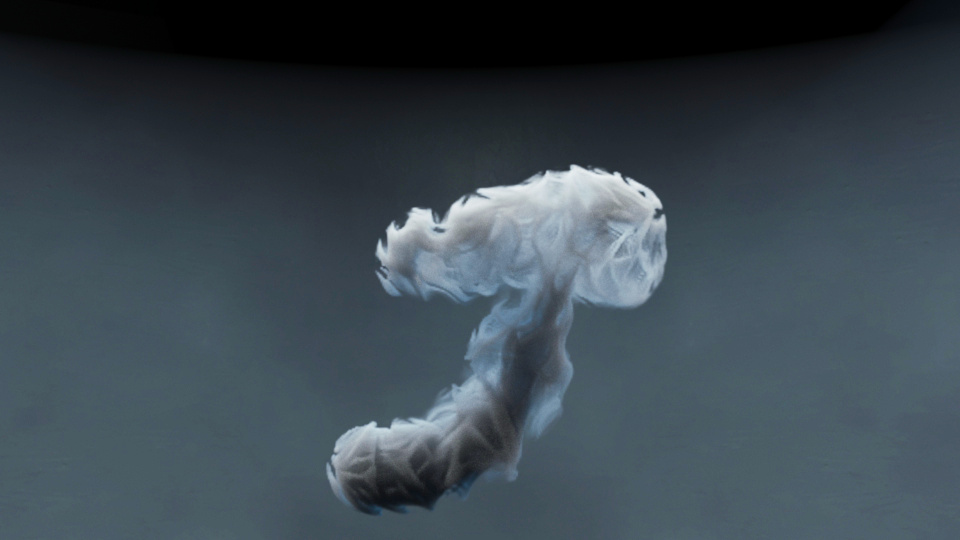}
    \hspace{-27.7px}\includegraphics[width=0.05\linewidth]{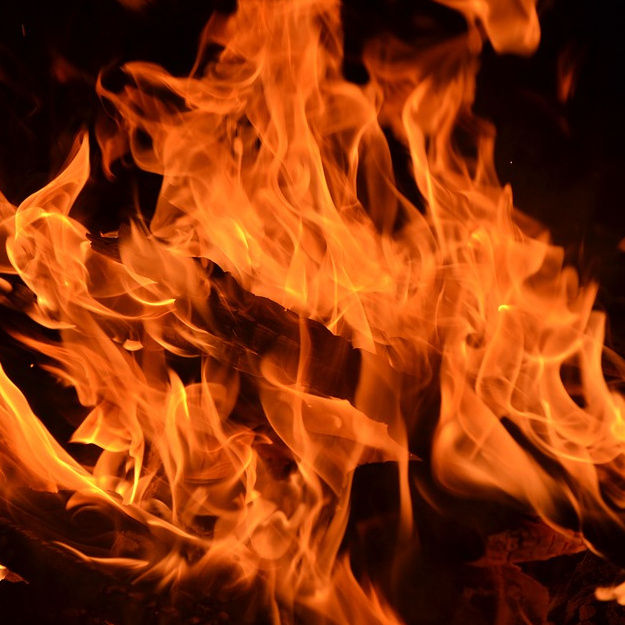}
    \includegraphics[trim=290px 0px 250px 110px, clip, width=0.15\textwidth]{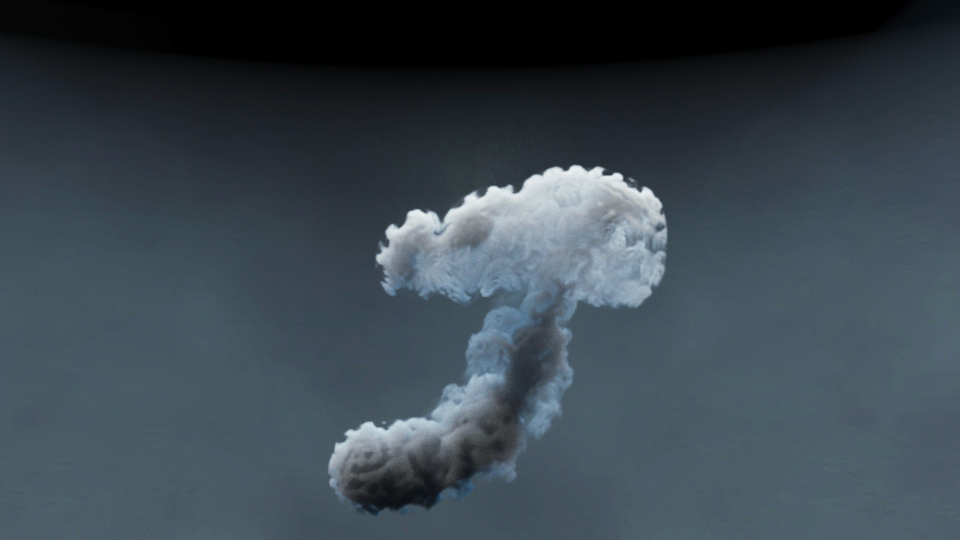}
    \hspace{-27.7px}\includegraphics[width=0.05\linewidth]{img/style/volcano}
    \includegraphics[trim=290px 0px 250px 110px, clip, width=0.15\textwidth]{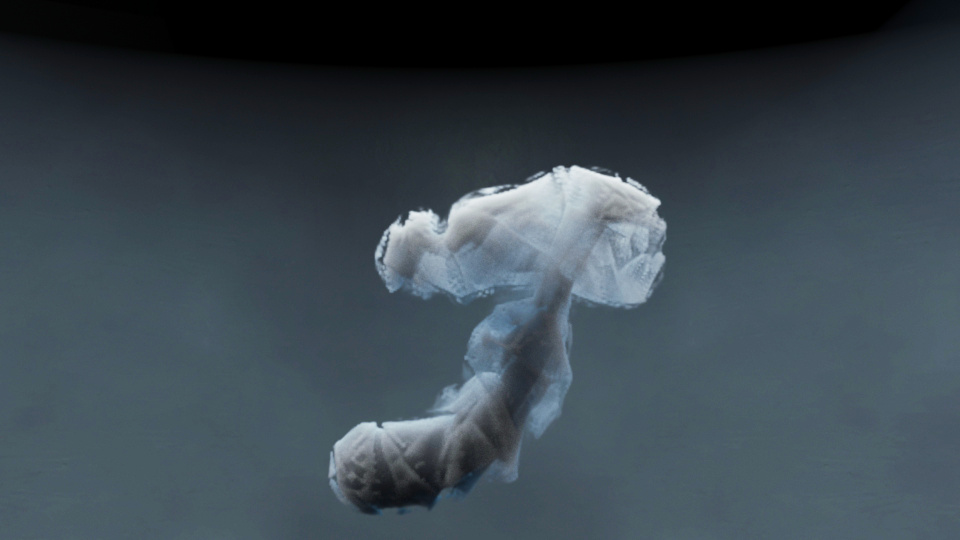}
    \hspace{-27.7px}\includegraphics[width=0.05\linewidth]{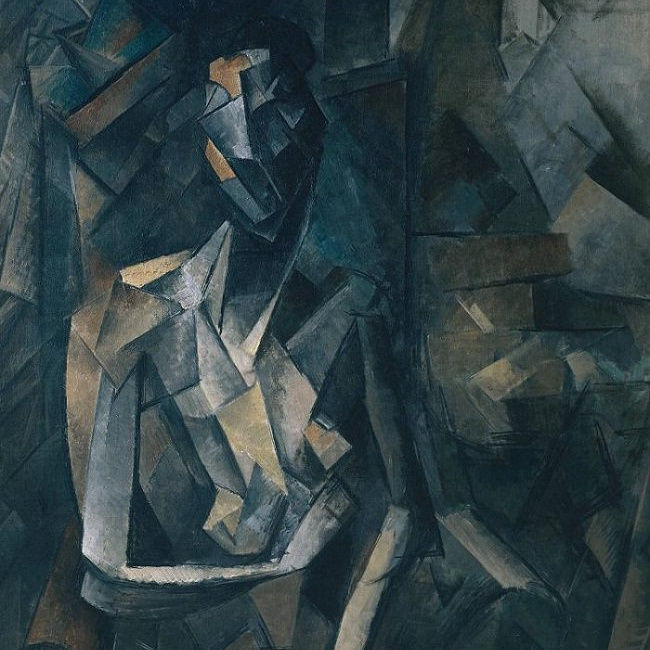}
    \includegraphics[trim=290px 0px 250px 110px, clip, width=0.15\textwidth]{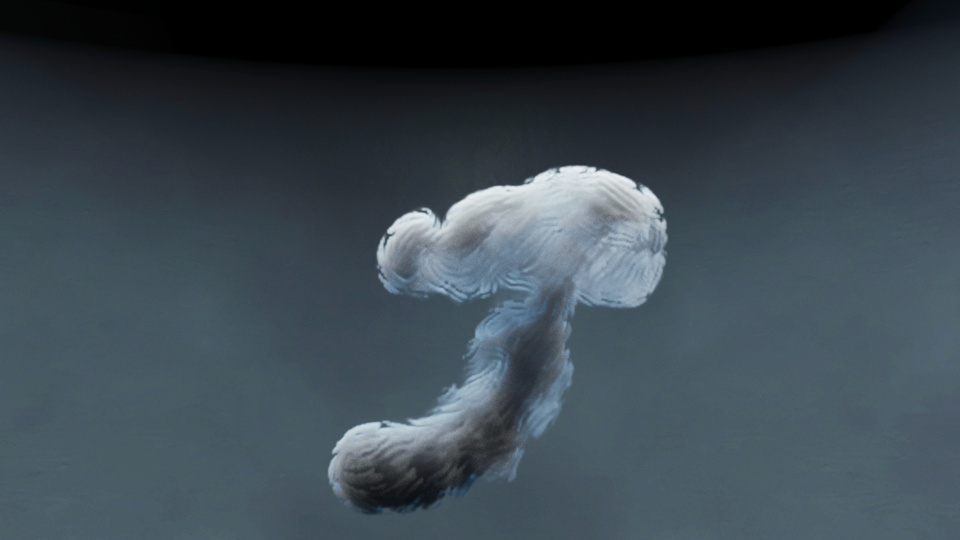}
    \hspace{-27.7px}\includegraphics[width=0.05\linewidth]{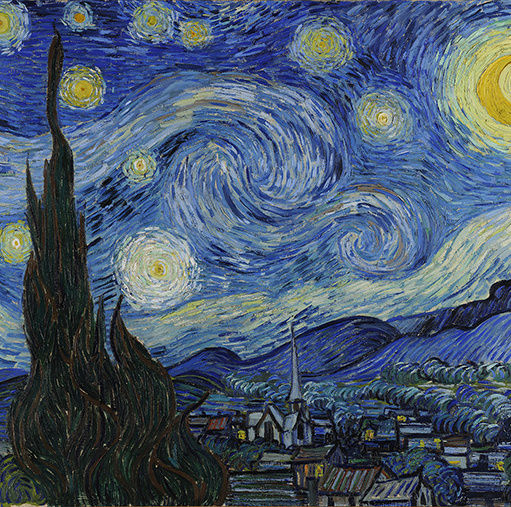}
    \includegraphics[trim=290px 0px 250px 110px, clip, width=0.15\textwidth]{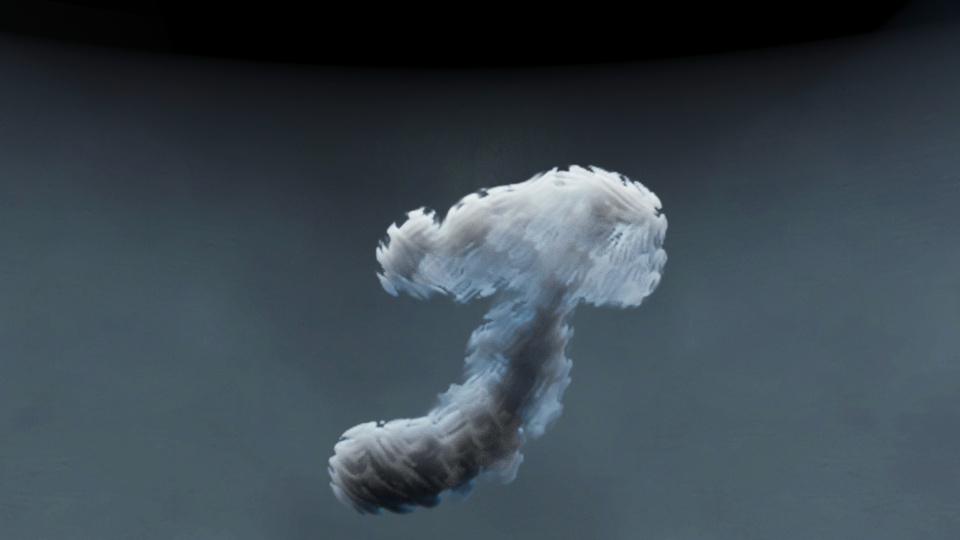}
    \hspace{-27.7px}\includegraphics[width=0.05\linewidth]{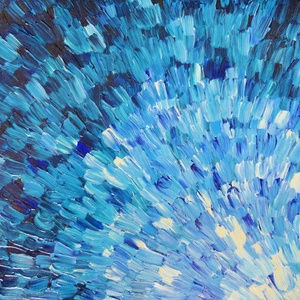}
    \includegraphics[trim=290px 0px 250px 110px, clip, width=0.15\textwidth]{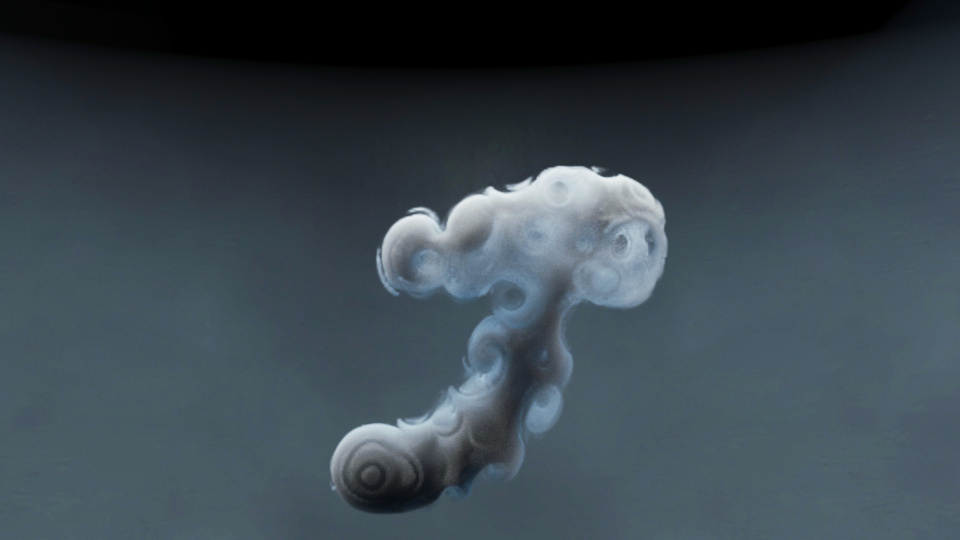}
    \hspace{-27.7px}\includegraphics[width=0.05\linewidth]{img/style/pattern1}
    \\
     \vspace{1pt}
    \includegraphics[trim=320px 0px 320px 250px, clip, width=0.15\textwidth]{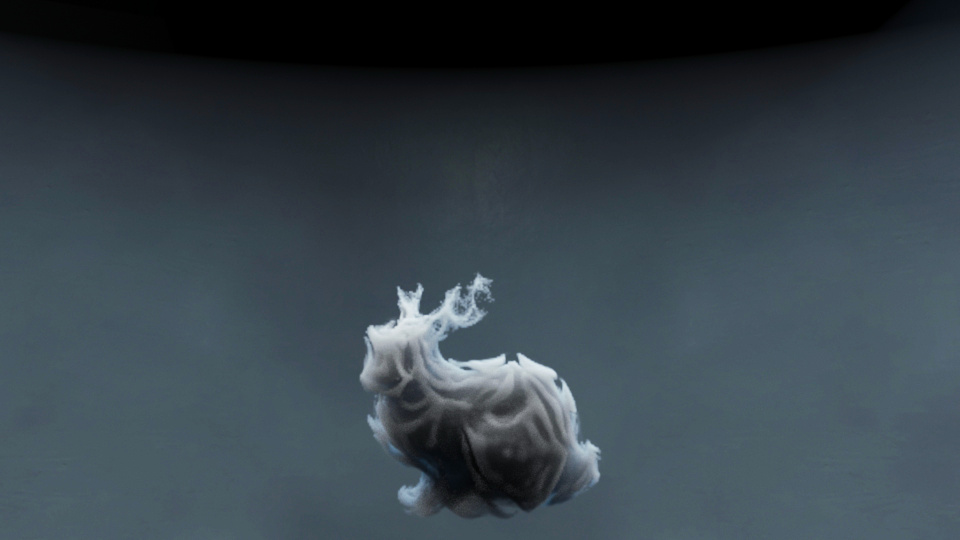}
    \includegraphics[trim=320px 0px 320px 250px, clip, width=0.15\textwidth]{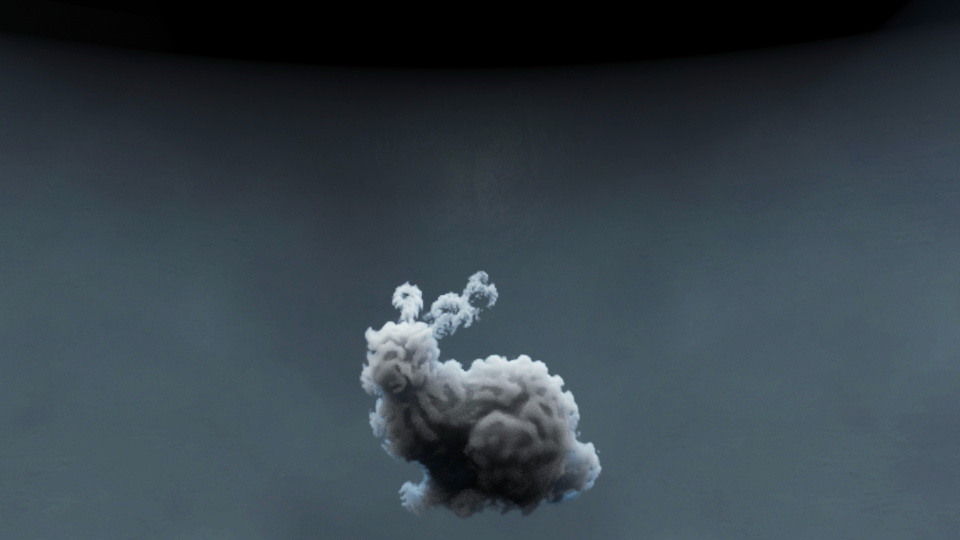}
    \includegraphics[trim=320px 0px 320px 250px, clip, width=0.15\textwidth]{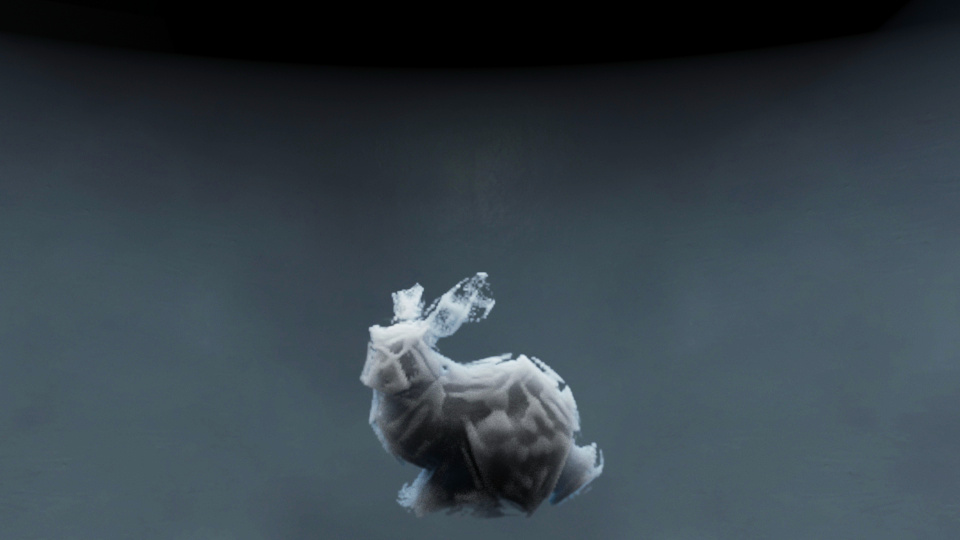}
    \includegraphics[trim=320px 0px 320px 250px, clip, width=0.15\textwidth]{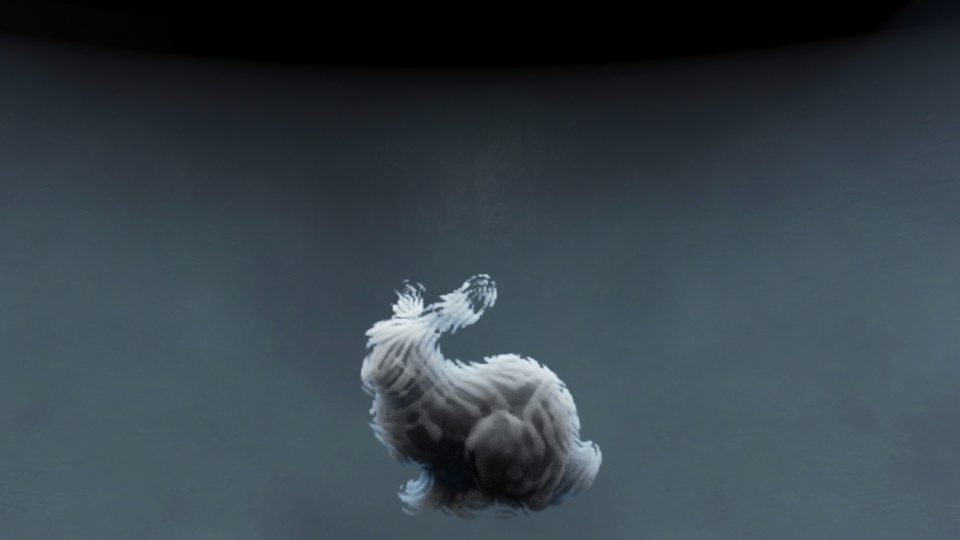}
    \includegraphics[trim=320px 0px 320px 250px, clip, width=0.15\textwidth]{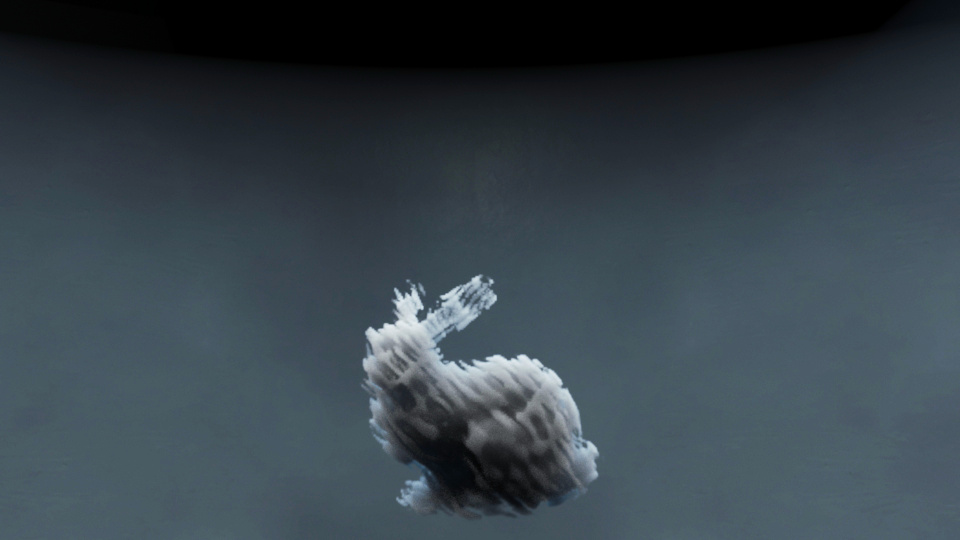}
    \includegraphics[trim=320px 0px 320px 250px, clip, width=0.15\textwidth]{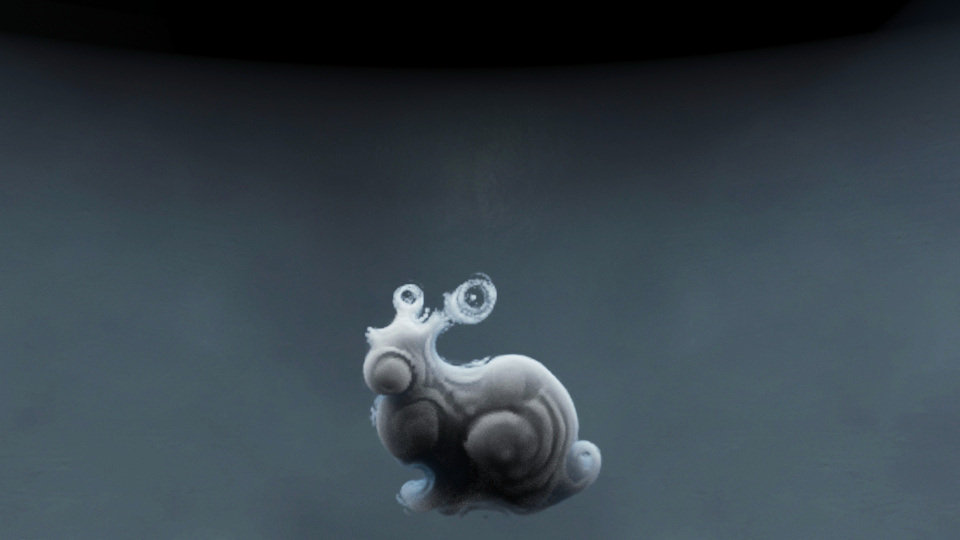}
    \caption{
Style transfer applied to a smokejet and bunny simulations. We used photorealistic (first two columns), artistic (middle two columns) and pattern-based (last two columns) input images\protect\footnotemark~as input to the stylization algorithm. Fire examplar \copyright Bunzellisa via pixabay.}
\label{fig:styleTransfer}
\end{figure*}

To demonstrate how our method performs under distinct style and semantic transfers, we designed two instances of buoyancy-driven smoke: a smokejet with a sphere-shaped source and an initial horizontal velocity, and a smoke initialized with the Stanford bunny shape (\Fig{semanticTransfer}, left). For all examples shown in Figures~\ref{fig:semanticTransfer} and \ref{fig:styleTransfer} we used 20 iterations for each scale with a learning rate of 0.002, 3 Laplacian subdivisions and 9 camera views for a single frame Poisson sampled around the original view with elevation ($\theta_1$) and azimuth ($\theta_2$) ranges spanning $[-5^\circ, 5^\circ]$ and $[-10^\circ, 10^\circ]$ respectively. The stylized examples show that our method is able to augment the original flow structures of the smoke, generating a wide set of artistic and natural \threeD~effects.

Examples in \Fig{semanticTransfer} demonstrate results obtained by applying the semantic style transfer loss from \Eq{semanticLoss}. All the feature maps are from the GoogleNet \cite{Szegedy2014} architecture; the captions indicate which layer of the CNN is used for the optimization. Activating different layers of the network will yield stylization results which will vary in the complexity of instantiated structures. In the two first examples we used filters closer to initial layers, which depict simpler patterns that occur at lower levels of abstraction. These patterns are used by higher levels to composite more complex structures. As the layers become deeper, the network is able to produce more intricate motifs, such as structures similar to flowers, fur, or ribbons.

\begin{figure}[h]\centering
   	\includegraphics[trim=350px 30px 310px 80px, clip, width=0.3\linewidth]{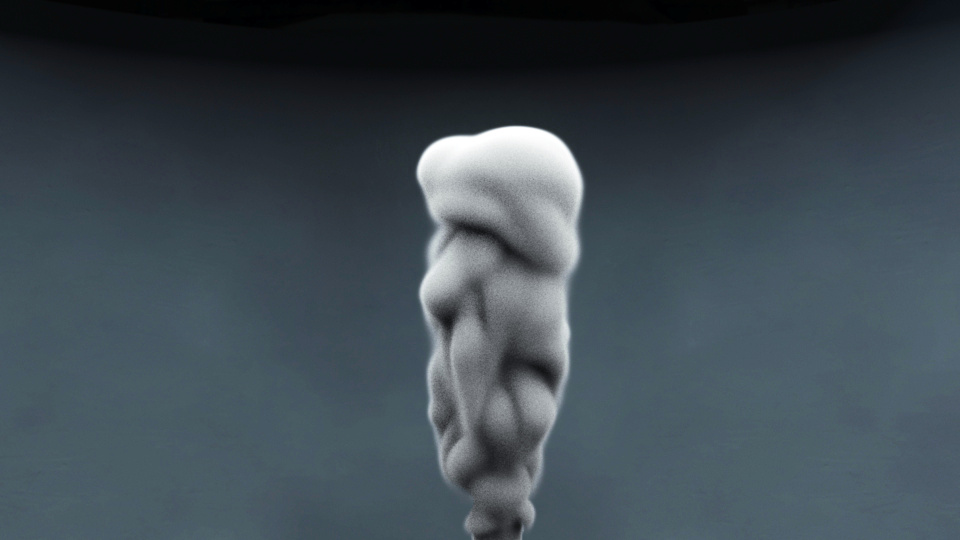}
    \includegraphics[trim=350px 30px 310px 80px, clip, width=0.3\linewidth]{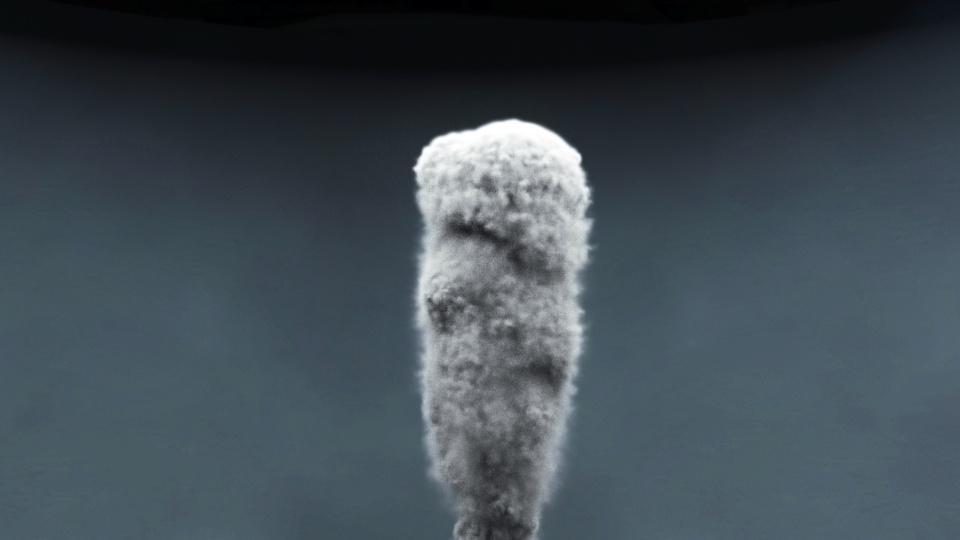}
    \includegraphics[trim=350px 30px 310px 80px, clip, width=0.3\linewidth]{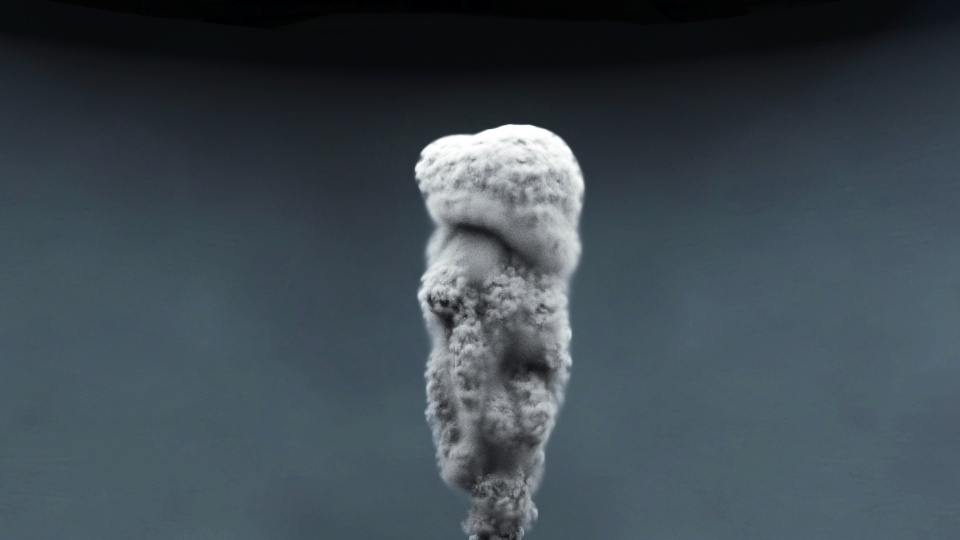}
    \caption{Style Transfer comparison. From left to right: input density field of Sato et al.~\shortcite{sato2018example}, the result of applying the method of Sato et al., and our style transfer result using the middle image as stylization input.
    }
	\label{fig:satoTransfer}
    \vspace{-10pt}
\end{figure}

Examples shown in \Fig{styleTransfer} apply the style loss of \Eq{styleTransferFinal}. To demonstrate the flexibility of our approach, we used three different image categories for testing the style transfer loss: photorealistic (first and second columns), artistic (third and fourth columns) and patterns (fifth and sixth columns). For all these examples, a mix of convolution layers from different levels of the CNN is used, similarly to \cite{gatys2015neural}. The representations of the layers employed on the style transfer are depicted below each input style image. \Fig{sequence} shows distinct frames of the bunny-shaped rising smoke stylized with the volcano and spiral input images from \Fig{styleTransfer}. In \Fig{satoTransfer} we compare our results with the example-based turbulence transfer method of Sato et al. \shortcite{sato2018example}. The style of the single frame rendering (\Fig{satoTransfer}, middle) of their method output is transferred by our approach to an input coarse simulation. The results show that our approach is able to generate similarly detailed flow structures with only a given reference image.

Besides individually transferring semantics and styles, our method is also flexible to allow the combination of these techniques. An example in the accompanying video demonstrates the semantic instantiation of cloud motifs merged with the style of a fire texture, while a similar stylization shows ribbon patterns combined with Starry Night painting style. \Fig{teaser} shows a low-resolution volcanic setup in which a combination of cloud motifs and a volcano texture was used to create turbulent details. The thick smoke produced by volcanic ashes poses challenges to our renderer, since it quickly saturates transmittance values. Nevertheless, our renderer is able to create structures that consistently correlate with the input smoke.

Our method is also able to control the amount of smoke dissipation by the decomposition of the stylization velocity field into its incompressible and irrotational parts. \Fig{fieldComparison} and examples in the supplemental video show that different artistic patterns are created depending on the constraints imposed on the velocity field.

\footnotetext{\url{https://github.com/titu1994/Neural-Style-Transfer}}

\begin{table*}
\caption{Parameters and performance statistics. We used a constant multi-scaling factor of 1.8, and the input size is firstly down-sampled to $61\times92\times61$ and up-scaled to $111\times166\times111$ and $200\times300\times200$. Computation time per frame includes all input scales.}
\vspace{-5pt}
\begin{tabular}{p{3.3cm} cccccccc}
\hline
& \small{Simulation} & & \small{Learning} & \small{Extinction} & \small{Multi} & \small{\# Target} & \small{Computation} \\
\small{Scene} & \small{Resolution} & \# \small{Frames} & \small{Rate} & \small{Factor $\alpha$} & \small{Scale} & \small{Layers} & \small{Time per Frame (m)}
%
\tabularnewline \hline
\small{Semantic Transfer} \footnotesize{(Fig.~\ref{fig:semanticTransfer})} & $200\times300\times200$ & 120 & 0.002 & 0.1 & 3 & 1 & 13.47\\
\small{Style Transfer} \footnotesize{(Fig.~\ref{fig:styleTransfer})} & $200\times300\times200$ & 120 & 0.002 & 0.1 & 2 & 3 & 12.68\\
\small{Volcano} \footnotesize{(Fig.~\ref{fig:teaser})} & $200\times300\times200$ & 140 & 0.003 & 10 & 2 & 2 & 12.97\\
\small{Sato et al.~\shortcite{sato2018example}} \footnotesize{(Fig.~\ref{fig:satoTransfer})} & $192\times256\times192$ & 140 & 0.005 & 5 & 2 & 3 & 11.97
\tabularnewline \hline
\end{tabular}
\label{tab:stat}
\vspace{-5pt}
\end{table*}

Although our method is based on \twoD~representations of the smoke data, it can reliably cover multiple viewing directions without introducing bias towards certain views. This is allowed by the Poisson sampling of positions along the camera trajectory. \Fig{multiView} shows the bunny smoke example stylized with a spiral pattern from different viewpoints. Transferred structures change smoothly when the camera moves around the object. An example in the accompanying video compares results of the smoke bunny example stylized with single view and multiple camera views. The multiple camera views are Poisson sampled with $180$ degrees angle range around the $y-$axis. The multi view approach shows more salient 3D structures when compared with the single view stylization.

Lastly, \Fig{temporalCoherenceWindowSize} shows the impact of different time coherency window sizes (\Eq{transport}). We compare a window size of 1 (frame-based stylization) with a window size of 9 (used in all other examples) for multiple subsequent frames. Features heavily flicker with a small window size, while augmented structures change smoothly with larger windows. These results are better visualized in the accompanying video, in which we also included a comparison with a window size of 5.

\subsection{Discussion}
\label{sec:discussion}
\paragraph{Differentiability} Note that all operators including advection and rendering require differentiability with respect to the velocities, so efficient gradient-based optimization methods can be employed. Traditional NST works by differentiating the loss of a classification network with respect to the image input, computing gradients of filter responses to image variations. Since classification networks convolve images to create filter responses, these filters are assumed to smoothly change with respect to image variations, and thus NST works without differentiability issues. This is the same for our work, however we additionally require that both the transport towards stylization and the smoke rendering to be differentiable. The rendering scheme denoted by \Eq{renderingEquation} is clearly differentiable. The MacCormack advection uses the Semi-Lagrangian method as its building blocks to correct error estimations. The correction is differentiable and the Semi-Lagrangian algorithm works by sampling densities in previous positions. Thus, two components need to be considered for differentiation to work: estimation of particle trajectories and density sampling. The estimation of the particle trajectories is a linear ODE, and thus differentiable. Densities are estimated by grid sampling, and as shown in \cite{Jaderberg2015} this is also differentiable when using linear interpolation kernels.

\paragraph{Performance and memory limitations} \Tab{stat} shows the average time for stylizing a single frame of different simulation resolutions, with grids up to $200 \times 300 \times 200$. Performance was not the focus of this work, and as shown by extensive follow-up works to image-based NST \cite{Ulyanov2016}, we believe that real-time stylizations can be obtained by training networks to directly output stylized results. For higher resolutions, the limiting factor is the single GPU memory used for computing the backpropagation with Tensorflow automatic differentiation. As in \cite{liu2018paparazzi}, the memory limitation could be greatly reduced by using analytic differentiation.

\paragraph{Temporal Coherency and Boundary conditions} Features are instantiated by evaluating smoke representations independently for each frame. Our temporal coherency algorithm aligns and blends the creation of those features; however, due to the nature of the underlying physical phenomena, smoke structures might appear and disappear as the simulation advances. This might induce abrupt changes in the stylized smoke, specially when considering smoke edges. We did not post-process our results in order to have a fair evaluation, but this effect can be controlled by blending the results with the original smoke simulation or by a more aggressive masking scheme. Regarding boundary conditions, the final stylized smoke can only slightly penetrate objects inside the simulation, since it starts from a density field configuration that is already boundary respecting. We include stylization experiments for scenes that include obstacles in our supplementary material. aterial.
 
\section{Conclusions}
In this work, we presented the first transport-based Neural Style Transfer algorithm for smoke simulations. Our method enables automatic instantiation of a vast set of motifs through semantic transfer, which enables novel artistic manipulations for fluid simulation data. Additionally, the proposed method successfully synthesizes various different styles of input images, ranging from artistic to photorealistic examples. Even though \twoD~CNNs are employed, our differentiable renderer allows the creation of \threeD~volumetric structures from small set of views. Our stylization algorithm is able to handle high resolutions simulations up to 16 million voxels.


We are not aware of any other methods that use a volumetric differentiable rendering for optimizing \threeD~smoke data, and we believe that our work may inspire further research in this direction. For example, our differentiable renderer could be employed for reconstructing \threeD~smoke volumes from images as in \cite{Eckert2018}, and be extended to transfer image-based filters as in \cite{liu2018paparazzi}. As further extensions, super-resolution can be thought as a specific type of style transfer \cite{Johnson2016Perceptual}, and we believe that our work can be tailored towards improving current super-resolution methods for fluids \cite{xie2018tempogan}. Additionally, the style transfer quality could be improved by histogram normalization \cite{Risser2017}. Our method does currently not handle color information to enable appearance transfer effects as shown in \cite{jamrivska2015lazyfluids}, which could be an interesting direction for further research.




\begin{acks}
The authors would like to thank Fraser Rothnie for his artistic contributions. The work was supported by the Swiss National Science Foundation under Grant No.: 200021\_168997.
\end{acks}

\bibliographystyle{ACM-Reference-Format}
\bibliography{flowstyle}
\end{document}